\documentclass[aps,prd,groupedaddress]{revtex4-1}

\usepackage{mathrsfs}
\usepackage{amsmath}
\usepackage{amssymb}
\usepackage{subfigure}

\usepackage{tikz}
\usetikzlibrary{decorations.pathmorphing}

\newcommand{\eps}{\varepsilon}
\newcommand{\del}{\partial}
\renewcommand*{\c}[1]{\mathbf{#1}}
\newcommand*{\ul}[1]{\underline{#1}}

\newcommand*{\tg}{\widetilde g}
\newcommand*{\tR}{\widetilde R}
\newcommand*{\hR}{\widehat R}
\newcommand*{\tP}{\widetilde P}
\newcommand*{\hP}{\widehat P}
\newcommand*{\tnabla}{\widetilde\nabla}
\newcommand*{\hnabla}{\widehat\nabla}
\newcommand*{\hGamma}{\widehat\Gamma}

\newcommand*{\hGammae}[3]{{\widehat\Gamma_{\c{#1}\c{#2}}{}^{\c{#3}}}}
\newcommand*{\hdal}{{\Box\kern-7.5pt\wedge}}

\newcommand*{\tK}{\widetilde K}
\newcommand*{\tU}{\widetilde U}
\newcommand*{\tgam}{\widetilde \gamma}
\newcommand*{\hgam}{\widetilde \gamma}
\newcommand*{\hG}{\widehat G}

\newcommand*{\Y}[2]{{}_{#1}\!\mathrm{Y}_{#2}}
\renewcommand*{\c}[1]{\mathbf{#1}}
\newcommand{\dd}{\mathrm{d}}

\newcommand*{\scri}{\mathscr{I}}

\newcommand*{\Mb}{\mathbf{M}}
\newcommand*{\ii}{\mathrm{i}}
\newcommand*{\ee}{\mathrm{e}}
\newcommand{\tpsi}{\widetilde\psi}

\allowdisplaybreaks

\begin{document}

% Use the \preprint command to place your local institutional report
% number in the upper righthand corner of the title page in preprint mode.
% Multiple \preprint commands are allowed.
% Use the 'preprintnumbers' class option to override journal defaults
% to display numbers if necessary
%\preprint{}

%Title of paper
\title{The numerical initial boundary value problem for the generalized conformal field equations}

% repeat the \author .. \affiliation  etc. as needed
% \email, \thanks, \homepage, \altaffiliation all apply to the current
% author. Explanatory text should go in the []'s, actual e-mail
% address or url should go in the {}'s for \email and \homepage.
% Please use the appropriate macro foreach each type of information

% \affiliation command applies to all authors since the last
% \affiliation command. The \affiliation command should follow the
% other information
% \affiliation can be followed by \email, \homepage, \thanks as well.

\author{Florian Beyer}
\email[]{fbeyer@maths.otago.ac.nz}
\affiliation{University of Otago, Dunedin, New Zealand}

\author{J{\"o}rg Frauendiener}
\email[]{joergf@maths.otago.ac.nz}
\affiliation{University of Otago, Dunedin, New Zealand}

\author{Chris Stevens}
\email[]{c.stevens@ru.ac.za}
\altaffiliation[Current address: ]{Rhodes University, Grahamstown, South Africa}
\affiliation{University of Otago, Dunedin, New Zealand}

\author{Ben Whale}
\email[]{bwhale@maths.otago.ac.nz}
\affiliation{University of Otago, Dunedin, New Zealand}

%Collaboration name if desired (requires use of superscriptaddress
%option in \documentclass). \noaffiliation is required (may also be
%used with the \author command).
%\collaboration can be followed by \email, \homepage, \thanks as well.
%\collaboration{}
%\noaffiliation

\date{\today}

\begin{abstract}
% In this paper we derive Friedrich's generalized conformal field equations in the space-spinor formalism and conduct the first numerical investigations into the associated initial boundary value problem. First, the system and numerical methods are thoroughly tested by solving several initial value problems, after which we present a method of fixing boundary conditions for the initial boundary value problem so that the constraints stay satisfied on at least the numerical level. This framework is then verified numerically using a simple case of gravitational perturbations of Minkowski space-time and subsequently with gravitational perturbations of Schwarzschild space-time.

  In this paper we study a numerical implementation for the initial boundary value formulation for the generalized conformal field equations. We propose a formulation which is well suited for the study of the long-time behaviour of perturbed exact solutions such as a Schwarzschild or even a Kerr black hole. We describe the derivation of the implemented equations which we give in terms of the space-spinor formalism. We discuss the conformal Gauss gauge, and a slight generalization thereof which seems to be particularly useful in the presence of boundaries. We discuss the structure of the equations at the boundary and propose a method for imposing boundary conditions which allow the correct number of degrees of freedom to be freely specified while still preserving the constraints. We show that this implementation yields a numerically well-posed system by testing it on a simple case of gravitational perturbations of Minkowski space-time and subsequently with gravitational perturbations of Schwarzschild space-time.
\end{abstract}

% insert suggested PACS numbers in braces on next line
% insert suggested keywords - APS authors don't need to do this
%\keywords{}

%\maketitle must follow title, authors, abstract, \pacs, and \keywords
\maketitle

% body of paper here - Use proper section commands
% References should be done using the \cite, \ref, and \label commands

\section{\label{sec:intro}Introduction}

% Introduce tensorial GCFE and issues with IBVP. No numerical investigations etc. 

In the 1960's Penrose introduced the idea of studying the asymptotic properties of a space-time from the perspective of its conformal (light cone) structure \cite{penrose1963asymptotic,penrose1962light,penrose1964conformal}. The idea is to conformally embed the physical space-time in question into a conformally related space-time, usually called the conformal space-time, where its image becomes an open submanifold. The boundary of this submanifold in the conformal space-time is referred to as the \emph{conformal boundary}, (denoted $\mathscr{I}$), and represents the points at infinity of the physical space-time. A very nice property of this picture is that from the point of view of the conformal space-time, the conformal boundary, and hence the asymptotic structure of the physical space-time, can be investigated using local differential geometry.

Although Penrose constructed the conformal completion of many  known space-times  the question as to whether there exist large classes of space-times which admit a conformal completion was not resolved over a long period of time. Over the years Friedrich has worked on this question by developing the \emph{conformal field equations}, which are the mathematical extension of the Einstein equations to the conformal manifold. The Generalized Conformal Field Equations (GCFE) are the latest form of these equations, and were first written down in 1995 in the important work \cite{friedrich1995einstein}. It was only after the surprising result by Corvino and Schoen~\cite{Corvino:2006wf,Corvino:2000vu} on the construction of initial data which agree with exact Schwarzschild or Kerr data outside a compact region that the question of existence raised above could be answered in the affirmative by Chru\'sciel and Delay~\cite{Chrusciel:2002vb}.

% In Einstein's general theory of relativity there is no covariant way to locally define energy. This is mainly to the problem of how to characterize the energy arising from gravitational radiation. However on the conformal boundary the geometry separates from the fields, and hence we can define some sort of global energy. Thus the conformal boundary becomes important to characterize asymptotic and global properties of space-times.

With the advent of gravitational wave detectors and, even more so, after the successful detection of (so far) three gravitational wave events~\cite{abbott2016gw151226,abbott2016observation,Abbott:2017cl}, the calculation of asymptotic quantities has become a major task for the numerical relativity community. After all, the waveforms used by the detectors as templates to match to observational data are strictly defined only on part of the conformal boundary. %Thus accurate and fast methods to calculate these asymptotic quantities are needed.

In the numerical relativity community it is standard practice to calculate the waveforms in the so-called ``wave-zone'', where the space-time is almost flat and where a decent approximation can be made, and not on the conformal boundary itself (with the exception of, for example, characteristic extraction \cite{reisswig2010characteristic}). The Cauchy-perturbative approach and extrapolation method are among the standard ways of calculating the waveforms in the wave-zone (see~\cite{bishop2016extraction} and the references within for a comprehensive review of these methods). These rely on solving the full non-linear Einstein equations in the portion of the space-time that has the most dynamics (for example inside a box surrounding a binary black hole system) generally using the BSSN formulation \cite{shibata1995evolution,baumgarte1998numerical}, the Generalized Harmonic Formalism \cite{pretorius2005numerical} or the Z4 formalulations \cite{bona2003general,alic2012conformal}. These approaches usually solve the Einstein equations in the physical space-time in the form of an Initial Value Problem (IVP) or an Initial Boundary Value Problem (IBVP) with a time-like outer boundary that approaches future time-like infinity.

The main difference between the conformal field equations over the standard approaches is that they grant access to the \emph{entire} conformal boundary, while the other methods remain in the physical space-time or, in some cases, extend to a finite portion of null infinity. With the standard approaches it is principally impossible to explore the global structure of a space-time completely since one is always confined to a finite portion of the physical space-time. Thus, one could not, for example, study the behaviour of fields across space-like infinity using Friedrich's cylinder interpretation \cite{Friedrich:1998tc} (see the summary \ref{sec:summary} for more applications). Yet, even with its advantages, there have been comparatively few investigations into the conformal field equation's potential for numerically studying global properties of asymptotically simple space-times, perhaps due to their apparent size and complexity. 

An early version of these equations --- nowadays referred to as the metric conformal field equations --- was studied in several papers in~\cite{frauendiener1998numerical1,frauendiener1998numerical2,frauendiener2000numerical,Frauendiener:2002ux} and in~\cite{Hubner:2001vb,Hubner:1999hz,Hubner:1999vv,Hubner:1996tc,Hubner:1996tf}. It could be demonstrated that the conformal approach does indeed deliver on its promise: one could evolve the global space-time within a finite computational time on a finite grid and  determine the gravitational wave-forms with high accuracy. However the downside of this particular formulation was that the conformal factor is an unknown in the system, being evolved alongside other quantities, and hence the location of the conformal boundary is not known \emph{a priori}, even though it was possible to devise suitable ``$\scri$-freezing'' gauges. 

The only numerical implementations of the generalized conformal field equations to date are due to Zengino\u{g}lu \cite{zenginouglu2007conformal,zenginouglu2007numerical} who reproduced the Schwarzschild and Kerr space-times in spherical and axi-symmetric symmetry, and by Beyer \cite{beyer2007asymptotics,beyer2009spectral,beyer2008investigations,beyer2009non} who considered the associated Initial Value Problem (IVP) for investigating spatially compact space-times with a positive cosmological constant $\lambda$. 

In this paper we want to study a numerical implementation of the GCFE in the context of an initial boundary value problem. We consider this approach for several reasons. Any numerical formulation of an evolution problem is based on a finite computational domain so, quite generally, some sort of conditions will have to be imposed on its boundary. Even when the mathematically underlying problem does not need any boundary conditions --- this will be the case when the problem is based on spatially \emph{compact} manifolds or when the boundary is a total characteristic such as when considering the finite initial boundary value problem at space-like infinity~\cite{Friedrich:1998tc} --- the finite size of the computational domain will lead to a numerical IBVP in the sense that the boundary points will need special consideration.

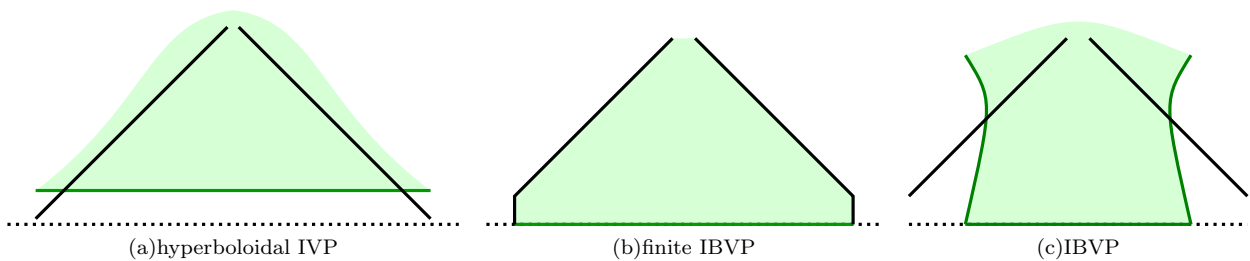
\begin{figure}[htb]
  \centering
  \subfigure[hyperboloidal IVP]{
    \begin{tikzpicture}[very thick,scale=0.75]
%  \draw[help lines,step=0.5] (-4,-1) grid (4,4);
  \fill[green!20, fill opacity=0.8]
       (-3.5,0.5)
           .. controls (-1.5,2) and (-1.5,3.5) ..
       (0,3.7) 
           .. controls (1.5,3.5) and (1.5,2) ..
       (3.5,0.5) -- cycle  ;
  \draw[green!60!black, very thick] (-3.5,0.5) -- (3.5,0.5);
  \draw[solid] (-3.5,0) -- (-0.1,3.4) (0.1,3.4) -- (3.5,0);
  \draw[dotted] (-4,-0.1) -- (4,-0.1);
\end{tikzpicture}
}
\subfigure[finite IBVP]{
\begin{tikzpicture}[very thick,scale=0.75]
%  \draw[help lines,step=0.5] (-4,-1) grid (4,4);
  \fill[green!20, fill opacity=0.8]
        (-3,0) -- (-3,0.5) -- (-0.2,3.3) -- (0.2,3.3) -- (3,0.5) -- (3,0) -- cycle;
  \draw[dotted] (-3.5,0) -- (3.5,0);
  \draw[green!60!black, very thick] (-3,0) -- (3,0);
  \draw[solid, very thick]
        (-3,0) -- (-3,0.5) -- (-0.2,3.3) (0.2,3.3) -- (3,0.5) -- (3,0);
\end{tikzpicture}
}
\subfigure[IBVP]{
\begin{tikzpicture}[very thick,scale=0.75]
%  \draw[help lines,step=0.5] (-4,-1) grid (4,4);
  \fill[green!20, fill opacity=0.8]
         (-2,0) .. controls (-1.5,2.2) .. (-2,3.0)  .. controls (0,3.8) .. (2,3.0) .. controls
                (1.5,2.2) .. (2,0) -- cycle;
  \draw[dotted] (-3,0) -- (3,0);
  \draw[green!50!black, very thick] (-2,0) -- (2,0);
  \draw[green!50!black, very thick]
         (-2,0) .. controls (-1.5,2.2) .. (-2,3.0)   (2,3.0) .. controls
                (1.5,2.2) .. (2,0);
  \draw[solid, very thick]
         (-3,0.5) -- (-0.2,3.3) (0.2,3.3) -- (3,0.5) ;
\end{tikzpicture}
}
\caption{\label{fig:setups} Schematic diagrams showing the different setups used in the context of the conformal field equations for the $\lambda=0$ case. The black solid lines are null-infinity and the time-like cylinder at space-like infinity in case (b). The dotted horizontal line indicates an asymptotically Euclidean hyper-surface and the shaded region is the computational domain with the thick part of the boundary indicating where (physical) data are given, initial data in all cases and boundary data in (c).}
\end{figure}

Our approach complements the more traditional hyperboloidal initial value problem pursued in the context of the metric conformal field equations and the finite initial boundary value problem just mentioned. The hyperboloidal IVP is based on space-like hyper-surfaces which extend to null-infinity. Initial data are provided by solving the constraint equations associated with the conformal field equations. By construction the hyperboloidal IVP is semi-global in its nature in the sense that its solution covers the global future of the space-time but only a part of its past, see Fig.~\ref{fig:setups}.

The finite IBVP at space-like infinity is based on asymptotically Euclidean hyper-surfaces which extend out to space-like infinity represented as a cylinder (see~\cite{beyer2012numerical,frauendiener2014fully,frauendiener2016fully,Doulis:2013wp,Doulis:2017kg} for some studies of this approach in very simplified situations). Initial data are obtained by solving the conformal constraint equations and the solution is global covering the entire past and future of the space-time evolving from the initial data, see Fig.~\ref{fig:setups}.

Our present approach was chosen for its very physical simplicity: we specify the initial data of a known solution exactly and perturb it by incoming radiation. More precisely, we specify initial data on a space-like hyper-surface with boundary which is not assumed to extend to infinity. Therefore, we need to specify boundary conditions in order to make the problem well-posed. If the problem is indeed well-posed (and in this paper we argue that it is at least numerically well-posed) then this setup allows us to study perturbations of an exact solution in a clear physical way. We choose initial data corresponding to the exact solution and specify the perturbation by boundary conditions. In this way we disentangle the object of the disturbance from the cause of the disturbance. Thus, we avoid the solution of the constraint equations which is known to introduce uncontrolled radiative degrees of freedom. Instead, we can specify in a very controlled way the incoming degrees of freedom on the time-like boundary. The solution obtained in this way is time-global, i.e. global in time but not in space in the sense that there is a region near space-like infinity which is not covered by the evolution of the initial data, see Fig.~\ref{fig:setups}. 

The GCFE can also be used to evolve asymptotically anti-de Sitter like space-times which naturally leads to an IBVP with boundary conditions to be specified on the conformal boundary~\cite{friedrich1995einstein}. Finally, an IBVP is advantageous for studying perturbations of an exact solution as we will do here: the initial conditions are provided by the exact solution and the perturbation is injected from the boundary.  Proceeding in this way avoids the necessity to solve the constraint equation for perturbed initial conditions. Hence the main focus of this work is to conduct the first numerical investigations into the viability of a well-posed IBVP framework for the GCFE where the constraints stay satisfied on at least the numerical level.

% The only research into the IBVP for the GCFE was conducted by Friedrich \cite{friedrich1995einstein} who presented a framework for the case of Anti-de Sitter space-time with the choice of the boundary being the geometrically special surface $\mathscr{I}$. This was a particular case where the ideas used in formulating the Friedrich-Nagy gauge \cite{friedrich1999initial}, a well-posed formulation of the IBVP for the physical Einstein equations, could be extended to the GCFE. He then modified the evolution system by adding constraints in such a way that it remained symmetric hyperbolic as well as the subsidiary (constraint propagation) system consisting only of Ordinary Differential Equations (ODEs), which avoids the problem of ingoing constraint violating modes. This process however is very specialized due the unique properties of having $\mathscr{I}$ as the boundary, and in general the subsidiary system will contain Partial Differential Equations (PDEs). Hence a major task to overcome when choosing a more general boundary is how to deal with the possible ingoing constraint violating modes.

The structure of this paper is as follows: In section \ref{sec:derivation} we summarize the derivation of the GCFE and introduce the conformal Gauss gauge. We describe how we impose Newman and Penrose's $\eth$-calculus~\cite{Newman:1966wt,Goldberg:1967tf,Eastwood:1982tb,penrose1986spinors} to obtain proper spin-weighted equations and end the section with our derivation of the GCFE in the space-spinor formalism. In section \ref{sec:ibvp} the field equation for the gravitational spinor is analysed and a numerical procedure for imposing constraint preserving boundary conditions is presented. We then test our system numerically using as applications: the IBVP for gravitational perturbations of Minkowski space-time in section \ref{subsec:MinkowskiIBVP} and the IBVP for gravitational perturbations of Schwarzschild space-time in section \ref{subsec:schwarzschildIBVP}. The paper concludes in section \ref{sec:summary} with a brief summary and discussion of future applications.

We use the conventions of Penrose and Rindler \cite{penrose1986spinors,penrose1988spinors} throughout. In particular, we use the metric signature $(+,-,-,-)$ and define associated with any torsion-free connection $\nabla$: the Riemann tensor $[\nabla_a,\nabla_b]\alpha_c=-R_{abc}{}^d\alpha_d$, the Ricci tensor $R_{ab}:=R_{acb}{}^c$, the Ricci scalar $R:=R_a{}^a$ and the Schouten tensor as $P_{ab}:=-\frac14R_{[ab]}-\frac12\Big{(}R_{(ab)}-\frac16Rg_{ab}\Big{)}$, where $R_{[ab]}=0$ if the connection is Levi-Civita. Then the vacuum Einstein field equations take the form $R_{ab}=\lambda g_{ab}$ with cosmological constant $\lambda$. We will use the convention of denoting quantities associated to the physical metric $\tg_{ab}$ with a $\;\widetilde{}\;$ and use bold Latin indices to denote frame indices $e^a_\c{i}$. Transvections of space-spinors with the spin-frame $\{o,\iota\}$ are written as $\alpha_0 := \alpha_Ao^A$ and $\alpha_1 := \alpha_A\iota^A$. If a space-spinor is symmetric in a certain number of indices, then we label their components as the number of contractions with $\iota$. For example if $K_{AB}=K_{(AB)}$ then its components are denoted by $K_0,\;K_1$ and $K_2$.

\section{\label{sec:derivation} The analytical background}

In this section we provide a brief derivation of the GCFE and the conformal Gauss gauge, a particularly useful set of gauge conditions for coordinates, tetrads and conformal factor. We discuss our imposition of the $\eth$-calculus and then use the space-spinor formalism to split the GCFE into evolution and constraint equation. Finally, we discuss the \emph{subsidiary system} which governs the violation of the constraints. For a more detailed derivation see for example \cite{friedrich2002conformal}.

\subsection{The general conformal field equations}
\label{sec:gener-conf-field}

Here we give a brief derivation of the GCFE, beginning with some mathematical preliminaries. The most important part of the GCFE is obtained from the Bianchi identities for the physical and the conformal metric together with Einstein's field equation which we take to be vacuum here, but admitting a non-vanishing cosmological constant $\lambda$. From these we obtain equations for the rescaled Weyl tensor and the Schouten (or, equivalently, the Ricci tensor) for a conformal Weyl connection. We write these equations in terms of a tetrad and obtain the equations for the tetrad components as well as the connection coefficients from Cartan's structure equations as usual.

We start with a vacuum space-time $(\widetilde{M},\tilde{g})$ where $\tilde{g}$ is a solution to the Einstein field equation with cosmological constant $\lambda$
\begin{equation*}
	\tilde{R}_{ab} = \lambda\tilde{g}_{ab},
\end{equation*}
with $\tilde{R}_{ab}$ denoting the Ricci tensor of $\tilde{g}$.

Let $g_{ab} = \Theta^2\tg_{ab}$ be a conformally related metric with conformal factor $\Theta$. We denote by $\nabla_a$ and $\tnabla_a$ the Levi-Civita connections of $g_{ab}$ and $\tg_{ab}$ respectively. Furthermore, let $\hnabla_a$ be a Weyl connection, i.e.\ a connection which is torsion free and compatible with the conformal class of $g_{ab}$ but not necessarily compatible with any metric in the conformal class. Then, there exist smooth 1-forms $f_a$ and $b_a$ such that
\begin{align}
	\hnabla_a\tg_{bc} &= -2b_a\,\tg_{bc},\\
	\hnabla_ag_{bc} &= -2f_a\,g_{bc}.\label{eq:1}
\end{align}
It follows immediately that the 1-forms are related via
\begin{equation}\label{eq:2}
	f_a = b_a - \Theta^{-1}\nabla_a\Theta.
\end{equation}
For reasons that will become apparent later, we also define another 1-form as
\begin{equation}\label{eq:11}
	h_a := \Theta b_a = f_a + \nabla_a\Theta.
\end{equation}
Any two covariant derivative operators differ by a $(2,1)$-tensor. For the difference between $\hnabla_a$ and $\nabla_a$ we denote this tensor by $f_{ab}{}^c$. Acting on co-vectors yields
\begin{equation}\label{eq:8}
	(\hnabla_a - \nabla_a)\omega_b = -f_{ab}{}^c\omega_c,
\end{equation}
from which it can be shown that
\begin{equation}\label{eq:9}
	(\hnabla_a - \nabla_a)v^c = f_{ab}{}^cv^b,
\end{equation}
where $f_{ab}{}^c$ is given by
\begin{equation*}
	f_{ab}{}^c = \delta^c{}_af_b + \delta^c{}_bf_a - g_{ab}g^{cd}f_d.
\end{equation*}

We now introduce arbitrary coordinates $x^\mu$ and frame field $e^a_\c{a}$, which is orthonormal with respect to the conformal metric $g_{ab}$ and we define $\eta_\c{ab} := g_{ab}e^a_\c{a}e^b_\c{b} = \mathrm{diag}(+1,-1,-1,-1)$. We then have (with the obvious notation $\hnabla_\c{a} = e^a_\c{a}\nabla_a$)
\begin{equation*}
	c^\mu_{\c{a}} := e_{\c{a}}(x^\mu) = \widehat{\nabla}_{\c{a}}x^\mu,
	 \qquad \hnabla_{\c{a}}e_{\c{b}} = \hGamma_{\c{ab}}{}^{\c{c}}e_{\c{c}}.
\end{equation*}
By writing \eqref{eq:1} in the basis $(e_\c{a})_{\c{a}=0:3}$ and contracting with $e^b_\c{b}e^c_\c{c}$, we find
\[
	f_\c{a}=\frac14\hGamma_\c{ab}{}^\c{b}.
\]

We are now in a position to write down the field equations. The first two are Cartan's two structure equations~\cite{kobayashi1963foundations}, where the torsion-free equation is given by
\begin{equation*}
		\Big{[}\hnabla_a,\hnabla_b\Big{]}x^\mu=0,
\end{equation*}
and the curvature equation is given by
\begin{equation*}
	\Big{[}\hnabla_a,\hnabla_b\Big{]}e_\c{c}^d=\hR_{abc}{}^de_\c{c}^c.
\end{equation*}
Contracting these equations with $e^a_\c{a}e^b_\c{b}$ and using the decomposition of the Riemann tensor in terms of the Schouten and Weyl tensors (note that the Schouten tensor is not necessarily symmetric for a general Weyl connection)
\begin{equation*}
	\hR_{abc}{}^d = 2\Big{(}\delta^d{}_{[a}\hP_{b]c} - 
	\delta^d{}_c\hP_{[ab]} -
	g_{c[a}\hP_{b]}{}^d\Big{)} + 
	C_{abc}{}^d,
\end{equation*}
yields the two equations
\begin{align}
e_{\c{a}}(c_{\c{b}}^\mu) - e_{\c{b}}(c_{\c{a}}^\mu) &= \hGammae{a}{b}{c} c_{\c{c}}^\mu - \hGammae{b}{a}{c} c_{\c{c}}^\mu,\label{GCFEeq:1} \\[12pt]
  e_{\c{a}}(\hGammae{b}{c}{d}) - e_{\c{b}}(\hGammae{a}{c}{d}) &=
  \hGammae{a}{b}{e} \hGammae{e}{c}{d} - \hGammae{b}{a}{e}
  \hGammae{e}{c}{d} - \hGammae{b}{c}{e} \hGammae{a}{e}{d} +
  \hGammae{a}{c}{e} \hGammae{b}{e}{d} \nonumber \\ 
& \hskip4em +\; C_{\c{a}\c{b}\c{c}}{}^{\c{d}} -
  2\eta_{\c{c}[\c{a}}\hat P_{\c{b}]}{}^{\c{d}} +
  2\delta_{[\c{a}}{}^\c{d}\hat P_{\c{b}]\c{c}} - 2\hat
  P_{[\c{a}\c{b}]}\delta_{\c{c}}{}^{\c{d}}. \label{GCFEeq:2}
\end{align}

The next equation comes from the Bianchi identity for the vacuum metric $\tg_{ab}$,
\begin{equation*}
	\tnabla_{[e}\tR_{ab]c}{}^d = \tnabla_{[e}C_{ab]c}{}^d = 0.
\end{equation*}
Rewriting this in terms of the conformal connection and gravitational tensor $K_{abc}{}^d:=\Theta^{-1}C_{abc}{}^d$ gives the simple equation
\begin{equation}\label{GCFEeq:3}
	\nabla_e K_{abc}{}^e = 0.
\end{equation}
Written in terms of the Weyl connection, we find
\begin{equation*}
	\hnabla_e K_{abc}{}^e = f_e K_{abc}{}^e.
\end{equation*}
The final equation is obtained from the Bianchi identity for the Weyl connection
\begin{equation*}
 	\hnabla_{[e}\hR_{ab]c}{}^d = 0,
 \end{equation*} 
 which, after a bit of work using~\eqref{GCFEeq:3}, gives
\begin{equation}
	\hnabla_a \hP_{bc} - \hnabla_b \hP_{ac} 
	= \left(\hnabla_e\Theta + \Theta f_e\right) K_{abc}{}^e
	= h_eK_{abc}{}^e. \label{GCFEeq:4}
\end{equation}
Then the full set of conformal field equations expressed with a Weyl connection is given by \eqref{GCFEeq:1},\eqref{GCFEeq:2},\eqref{GCFEeq:3} and \eqref{GCFEeq:4}, where in~\eqref{GCFEeq:2} $C_{\c{a}\c{b}\c{c}}{}^{\c{d}}$ is replaced by $\Theta K_{\c{a}\c{b}\c{c}}{}^{\c{d}}$.  

These equations are invariant under several transformations. Since they are obtained from geometric differential equations they are invariant under arbitrary coordinate and tetrad transformations. However, they are also invariant under the simultaneous rescalings $\Theta \mapsto \Theta \phi$ and $g_{ab} \mapsto \phi^2 g_{ab}$ with some arbitrary scalar field $\phi$ and under the change of Weyl connection $\hnabla_a$ which amounts to the choice of a 1-form $b_a$. Thus, we will have to fix the freedom in these transformations in order to get determined systems. This means that we need to fix coordinates, the tetrad field, the conformal factor $\Theta$ and the 1-form $b_a$ or, with $\Theta$ fixed, of $h_a$.

A particularly useful gauge is the \emph{conformal Gauss gauge} (CGG) which is obtained from the conformal structure itself and which we discuss next.

\subsection{The conformal Gauss gauge}
\label{sec:conf-gauss-gauge}

The conformal Gauss gauge is fixed solely from the conformal structure of the space-time, making use of the associated conformal geodesics. These are curves governed by the equations
\begin{subequations}
\begin{align*}
	u^b\nabla_bu^a &= -2 (h_bu^b)\, u^a + (g_{cd}u^cu^d)g^{ab}h_b, \\
	u^b\nabla_bh_a &= (h_bu^b)\,h_a - \frac{1}{2}g_{ab}u^b (g^{cd}h_ch_d) - u^bP_{ba},
\end{align*}
\end{subequations}
for the tangent vector to the curves $u^a$ and a smooth 1-form $h_a$ along them. Here, $\nabla_a$ is any covariant derivative compatible with the conformal class $[\tg_{ab}]$ and $P_{ab}$ is its corresponding Schouten tensor. The metric $g_{ab}$ is any representative of the conformal class. Note, that the covariant derivative is not necessarily compatible with this metric. The equations are covariant under the change of Weyl connection in the following sense: let us suppose that we have a solution $\{u^a, h_a\}$ of the conformal geodesic equations above. Let $f_a$ be any smooth 1-form then $\{u^a, h_a - f_a\}$ is a solution of the conformal geodesic equation with $\nabla_a$ replaced by another Weyl connection $\hnabla_a$ which is defined by
\[
\hnabla_a g_{bc} = \nabla_a g_{bc} - 2 f_a g_{bc}.
\]
We can now introduce the CGG according to the following procedure: we pick an initial space-like hyper-surface~$\widetilde{\Sigma}_0$ in the physical vacuum space-time $(\widetilde{M},\tg)$. Then we can write the conformal geodesic equation in terms of the physical metric $\tg_{ab}$ and its associated Levi-Civita derivative $\tnabla_a$. Note, that the Schouten tensor in these equations will reduce to a term proportional to $\lambda u_a$. We now select initial data for these equations on $\widetilde{\Sigma}_0$. We denote quantities which are defined only on $\widetilde{\Sigma}_0$ by an underline. On $\widetilde{\Sigma}_0$, we choose a conformal factor $\ul{\Theta}$, a 1-form $\ul{b}_a$, coordinates $(\ul{x}^1, \ul{x}^2, \ul{x}^3)$, and a frame $\ul{e}_\c{a}$ which is orthonormal with respect to the conformal metric $\ul{g}_{ab}=\ul{\Theta}^2\ul{\tg}_{ab}$ and such that $\ul{e}_{\c{0}}$ is normal to~$\widetilde{\Sigma}_0$. Then we put $u^a:= e_\c{0}^a$ and use $(\ul{u}^a,\ul{b}_a)$ as the initial data for the conformal geodesic equations. Their solutions $(u^a,h_a)$ provide a unique time-like conformal geodesic through each point of $\widetilde{\Sigma}_0$ at least near $\widetilde{\Sigma}_0$. Provided  the initial data are sufficiently smooth, this congruence is smooth and caustic free in a local neighbourhood $U\subset \widetilde{M}$ of $\widetilde{\Sigma}_0$. Thus, at each point in $U$ we have a 1-form $h_a$. We denote this 1-form by $b_a$ and use it in place of $f_a$ to define a Weyl connection~$\hnabla_a$ as shown above.  Thus, the 1-form $b_a$ mediates between $\hnabla_a$ and the physical connection $\tnabla_a$. Written with respect to the new Weyl connection, the conformal geodesic equations simplify to
\begin{subequations}
\begin{align}\label{eq:3}
	u^b\hnabla_bu^a &=0,\\
	u^a\hP_{ab} &=0.\label{eq:uSchouten}
\end{align}
\end{subequations}
We can now use this congruence of time-like curves to define on $U$ the parameter $s$ as the (time) coordinate $x^0$ and the coordinates $(x^1,x^2,x^3)$ by the requirement that they be constant along the curves and agree with $(\ul{x}^1,\ul{x}^2,\ul{x}^3)$ on~$\widetilde{\Sigma}_0$. A tetrad field $e_\c{a}$ is defined on $U$ by parallel transport of the chosen frame on $\widetilde{\Sigma}_0$ using the Weyl connection $\hnabla_a$. Thus, each frame vector satisfies the equation
\begin{equation}
  \label{eq:tetradtransport}
  0 = \hnabla_ue^c_{\c{c}} = u^a\tnabla_a e^c_\c{c} + (b_ae^a_\c{c})\, u^b + (b_au^a)\, e^b_\c{c} - (\tg_{ab}u^ae^b_\c{c})\, \tg^{ec}b_e .
\end{equation}
We point out here that there is an additional freedom in the transport of the frame along the congruence. Instead of parallel transport we can rotate the spatial legs of the frame in an arbitrary way without affecting the geometry. This amounts to imposing the slightly more general equation 
\begin{equation}
  \label{eq:tetradtransportrotation}
 \hnabla_ue^c_{\c{c}} = \omega^{\c{b}}{}_{\c{c}} e^c_{\c{b}}, \qquad \text{for }\c{c} = 1,2,3
\end{equation}
where $\omega^{\c{b}}{}_{\c{c}}$ is an infinitesimal spatial rotation, i.e. it is characterized by
$\omega_{\c{bc}} = - \omega_{\c{cb}}$ and $\omega_{\c{0b}} = 0$. It can be arbitrarily prescribed and we will use this freedom partially below.

The tetrad field defines a metric $g_{ab}$ which is necessarily conformal to the physical metric $\tg_{ab}$ with a conformal factor $\Theta$ given by the equation
\[
g(e_\c{a},e_\c{b})=\eta_\c{ab} = \Theta^2 \tg(e_\c{a},e_\c{b}).
\]
Note, that these equations imply in particular that $\Theta^{-2} = \tg(u,u)$ from which we find using~\eqref{eq:1} and~\eqref{eq:3}
\begin{equation}\label{eq:4}
	\dot\Theta = (b_eu^e)\,\Theta,
\end{equation}
where we used the over-dot to denote the derivative along $u^a$. This equation can be used to propagate the conformal factor. Using this and the conformal geodesic equations, we could also find a propagation equation for the 1-form $h_a$ that appears in the GCFE, thus fixing the gauge completely. However, a closer look at these equations reveals a surprising consequence: when looking at successive derivatives of the conformal factor one finds
\begin{equation*}
	\dddot\Theta=0.
\end{equation*}
Thus, $\Theta$ is given as a quadratic polynomial in $s$. Its derivatives can be expressed in terms of the scalars 
\begin{equation*}
Z = \Theta  b_au^a = \dot\Theta, \quad H = \Theta^{-1}(\tg^{ab} b_ab_b - \frac13\lambda) = 2 \ddot\Theta,
\end{equation*}
and with these we obtain an explicit formula for the conformal factor, namely
\begin{equation}\label{eq:10}
	\Theta(s) = \ul{\Theta}  + \ul{Z} s + \frac14 \ul{H}\, s^2.
\end{equation}
in terms of initial data defined on $\widetilde{\Sigma}_0$.

Furthermore, the conformal geodesic equation and the fact that the frame is transported along the curve using~\eqref{eq:tetradtransportrotation} yield an equation for the \emph{frame components} of the 1-form $b_a$. With vanishing infinitesimal rotation $\omega^{\c{b}}{}_{\c{c}}$ and written with respect to the physical geometry it becomes 
\[
\frac{\dd}{\dd s}b_{\c{c}} = - (u^ab_a) b_{\c{c}} + \frac12 \Theta H\tg_{ab} u^a e_{\c{c}}^b.
\]
Since $u^a = e^a_\c{0}$ and with \eqref{eq:4} we can obtain an equation for the frame components of the 1-form $h_a = \Theta b_a$ appearing in the GCFE
\[
	\frac{\dd}{\dd s}h_{\c{c}} = \frac{1}2H \,\eta_{\c{0}\c{c}}.
\]
Noting that $\dot H = 2 \dddot\Theta =0$, we find the remarkable result that
\begin{equation}\label{eq:5}
  	h_{\c{0}}(s) = \frac{1}2 \ul{H}s + \ul{h}_{\c{0}},\qquad
  	h_{\c{c}}(s) = \ul{h}_{\c{c}},\qquad \c{c}=1,2,3.
\end{equation}
When the more general frame transport equation~\eqref{eq:tetradtransportrotation} with non-vanishing $\omega$ is imposed, then the equation for the spatial components of $h_a$ changes into
  \begin{equation}
\frac{\dd}{\dd s}h_{\c{c}} = \omega^{\c{b}}{}_{\c{c}} h_{\c{b}} \qquad  \c{c} = 1,2,3.\label{eq:13}
\end{equation}
Then, the equation cannot be solved explicitly anymore, but must be added as an evolution equation to the system.

In summary then, the conformal Gauss gauge is characterized by 
\begin{enumerate}
\item the coordinate conditions
  \[
    \hnabla_ux^\mu = \delta^\mu_0,
  \]
\item  the frame conditions $u^a = e_{\c{0}}^a$ and
  \[
    \hnabla_u e_{\c{b}} = \omega^{\c{c}}{}_{\c{b}}
    e_{\c{c}},
  \]
\item   the explicit form~\eqref{eq:10} for the conformal factor and 
\item the conditions for the conformal connection expressed either in the explicit
  form~\eqref{eq:5} or the transport equation~\eqref{eq:13} for $h_{\c{a}}$  and the
  condition on the Schouten tensor~\eqref{eq:uSchouten} which when
  expanded in the frame becomes
  \[
    \widehat{P}_{\c{0a}}=0.
  \]
\end{enumerate}
Let us point out a particular property of this gauge. The fact that the frame is parallel along the curves implies that, in general, it will not be aligned with the hyper-surfaces defined by constant $s$. This implies that the expansion of the tetrad vectors with respect to the coordinate basis will include a component along $\del_s$ and that the causal character of these hyper-surfaces may change during the evolution.

\subsection{\label{subsec:ethoperators}Imposing the $\eth$-calculus}

The GCFE will allow us to evolve data from the initial surface up to and beyond null-infinity. This is a hyper-surface which is time-like, null or space-like depending on the cosmological constant, but it is always topologically of the form $\mathbb{R} \times S^2$. This property suggests to set up numerical methods which are adapted to this spherical topology. However, this is not straightforward since the sphere cannot be covered by a single coordinate chart. Numerically, this leads to the so called \emph{pole problem}, the fact that the usual polar coordinates become singular on the poles. Another aspect of this problem appears when one uses a frame formalism, since there are no globally defined frames on the sphere.

These problems have been tackled by several researchers using interpolation between two charts~\cite{Gomez:1997jh}, the cubed sphere~\cite{Lehner:2005hc}, pseudo-spectral methods~\cite{Bonazzola:1999ji} and the eth-calculus~\cite{Bartnik:2000ch}. To our knowledge there is no Finite Element treatment of the sphere. Here, we elaborate on the $\eth$-calculus which we have implemented in our code. This has already been discussed extensively in previous papers~\cite{Beyer:2014bu,Beyer:2015ud} so we will keep the exposition here reasonably short.

The $\eth$-calculus can be regarded as a generalization of spectral methods on the circle $S^1$ based on Fourier series to the sphere $S^2$. It is based on harmonic analysis on $S^3$ and transferred to $S^2$ using the so called Hopf fibration $S^3\to S^2$. In the context of General Relativity it was first discussed by Newman and Penrose~\cite{Newman:1966wt}, see also~\cite{Goldberg:1967tf,dray1985relationship,Dray:1986vi}. There are two essential points: the first is the fact that components of tensors of different rank on $S^2$ can be expanded in terms of a complete system of ``functions'' on the sphere, called the spin-weighted spherical harmonics~$\Y{s}{lm}$. And the second point is that the covariant derivative in the direction of a certain complex vector is diagonal when written with respect to the~$\Y{s}{lm}$. 

To briefly discuss the fundamentals of the $\eth$-formalism we consider the unit sphere $S^2$ with its metric $q$ which we take to be negative definite here in order to agree with our conventions for the space-time metric. At each point of $S^2$ we can introduce an orthonormal frame or, equivalently, a complex tangent vector $\Mb$ with the properties $q(\Mb,\Mb)=0$ and $q(\Mb,\overline{\Mb}) = -1$. Obviously, this vector is not uniquely defined but only up to a phase transformation
\begin{equation}
  \label{eq:phasetrafo}
  \Mb \mapsto \ee^{\ii\alpha} \Mb,
\end{equation}
which corresponds to a frame rotation with the angle $\alpha$.

It is well known that the sphere does not admit a globally defined frame field, hence no globally defined vector $\Mb$. To reach every point on $S^2$ one needs to patch several charts with frames defined on them together in order to cover the entire sphere. On a point in the overlap region between two charts the frames respectively, the complex vectors from the two patches differ by a rotation, respectively a phase transformation of the form~\eqref{eq:phasetrafo}. Components of tensor fields on $S^2$ transform under a well defined way under this change of frame on the overlaps. As an example consider a 1-form $v$. Its contraction $\left<v,\Mb\right>$ with $\Mb$ transforms under~\eqref{eq:phasetrafo} according to $\left<v,\Mb\right> = \ee^{\ii\alpha} \left<v,\Mb\right>$. Components of higher rank tensors transform in similar characteristic ways under frame rotations. This leads to the notion of a spin-weighted ``function'' with spin-weight $s$ on $S^2$, i.e., a quantity $\eta$ which transforms as $\eta \mapsto \ee^{\ii \alpha s}\eta $ under~\eqref{eq:phasetrafo}. Such a quantity can be regarded as globally defined on $S^2$ since it essentially incorporates the collection of a component of a globally defined tensor field with respect to different charts \footnote{Strictly speaking, spin-weighted functions are global sections of an appropriate line bundle over $S^2$ associated with the frame bundle.}.

A spin-$s$ quantity $\eta$ can be represented as a series
\begin{equation}
  \label{eq:Ylmexp}
  \eta = \sum_{l=|s|}^\infty \sum_{m=-l}^l \eta_{lm} \,\Y{s}{lm}
\end{equation}
where the $\Y{s}{lm}$ are the spin-weighted spherical harmonics, a generalization of the well-known spherical harmonics $Y_{lm} = \Y{0}{lm}$ (see~\cite{penrose1986spinors} for more details).

With spin-weighted quantities being globally defined their derivatives should share this property. The usual coordinate derivatives or directional derivatives along frame vectors do not achieve this. Instead one uses the connection $\overset{q}{\nabla}$ associated with the metric $q$ to define covariant derivatives along $\Mb$ to obtain well-defined spin-weighted quantities. This leads to the definition of the $\eth$-operator: suppose $\eta = T(\Mb,\ldots,\overline{\Mb})$ is a spin-weighted quantity obtained from a tensor $T$ by evaluation on $k$ copies of $\Mb$ and $l$ copies of $\overline{\Mb}$ so that $\eta$ has spin-weight $s=k-l$. Then 
\begin{equation}
  \label{eq:ethdef}
  \eth\eta = \overset{q}{\nabla}_\Mb T(\Mb,\ldots,\overline{\Mb}).
\end{equation}
In similar way, the $\eth'$-operator is defined
\begin{equation}
  \label{eq:etpdef}
  \eth'\eta = \overset{q}{\nabla}_{\overline{\Mb}} T(\Mb,\ldots,\overline{\Mb}).
\end{equation}
Thus, $\eth$ raises the spin-weight by one, while $\eth'$ lowers it by one. It follows from these expressions that the action of $\eth$ and $\eth'$ on a spin-$s$ quantity $\eta$ can be alternatively expressed by
\begin{equation}
  \eth \eta = \Mb(\eta) - s \bar{\alpha}\,\eta,\qquad
  \eth'\eta = \overline{\Mb}(\eta) + s \alpha\,\eta,\label{eq:ethMb}
\end{equation}
where $\alpha$ is the single (complex) connection coefficient of $\overset{q}{\nabla}$ defined in terms of the commutator $[\Mb,\overline{\Mb}] = \alpha \Mb - \bar{\alpha}\overline{\Mb}$.

The action of these operators on the spin-weighted spherical harmonics is particularly simple
\begin{equation}
\label{eq:ethYlm}	
\eth\;\Y{s}{lm} = -\sqrt{(l-s)(l+s+1)}\;\Y{s+1}{lm}, \qquad	\eth' \Y{s}{lm} = \sqrt{(l+s)(l-s+1)}\;\Y{s-1}{lm}.
\end{equation}
It is this property which makes these operators useful for numerical purposes: to compute their action on any spin-weighted quantity $\eta = \sum_{l,m} \eta_{lm}\, \Y{s}{lm}$ is just a matter of multiplying the coefficients by the appropriate numbers and replacing $\Y{s}{lm}$ with $\Y{s\pm1}{lm}$.

To set up the framework for the $\eth$-calculus we need to make a few assumptions. As discussed in Sec.~\ref{sec:conf-gauss-gauge}, imposing the CGG fixes coordinates and tetrad in the space-time $M$ leaving us only the freedom to fix them on the initial hyper-surface $\Sigma_0$. We use this freedom to postulate that the initial hyper-surface $\Sigma_0$ is foliated by a family of 2-surfaces with spherical topology, so that it has topology $\mathbb{R}\times S^2$. We choose a radial coordinate $\rho$ on $\Sigma_0$ labelling the spherical leaves of this foliation. The remaining two coordinates label points on the spheres. 

On each sphere $S_\rho \subset\Sigma_0$ we introduce a (fiducial) negative definite unit-sphere metric $q = -( \dd \theta^2 + \sin^2\theta \, \dd \phi^2)$. This also introduces coordinates $\theta$ and $\phi$ which can be interpreted as the usual polar coordinates on each $S_\rho$. We assume that $\theta$ and $\phi$ are smooth on $\Sigma_0$ (except for the usual problems at the poles of the spheres). On each sphere $S_\rho$ we can also introduce a complex null-vector $\Mb$ as before.

The coordinates $(\rho,\theta,\phi)$ are constant along the conformal geodesics so that every event in the region of $M$ covered by the conformal Gauss gauge is characterized by coordinates $(s,\rho,\theta,\phi)$ with $s$ the parameter along the conformal geodesics. Thus, each event lies on a unique sphere $S_{s,\rho}$ defined by constant $s$ and $\rho$, carrying a negative definite unit-sphere metric $q$ and a complex vector $\Mb$.

We also need to pick a triad $(e_1,e_2,e_3)$ on $\Sigma_0$. We can arrange this to be adapted to the foliation in the sense that at every point $e_2$ and $e_3$ are tangent to the sphere $S_\rho$ through that point. Then $e_1$ is necessarily perpendicular to $S_\rho$. We can further arrange that $m:= \frac1{\sqrt2}(e_2 - \ii e_3)$ is proportional to $\Mb$ on $\Sigma_0$. The ambiguity in the choice of $\Mb$ is the same as the ambiguity in  the choice of an adapted triad, namely a phase transformation, respectively a frame rotation leaving $e_1$ fixed. 

As described in section~\ref{sec:conf-gauss-gauge}, the conformal Gauss gauge provides a space-time frame  $(e_0,e_1,e_2,e_3)$ on $M$ which it is orthonormal with respect to the conformal metric $g$. When the GCFE are referred to this frame they turn into equations for components with respect to the frame in terms of directional derivatives along the frame vector fields. The gauge conditions imply that the frame vectors can be written in terms of coordinate derivatives or, since $\del_\theta$ and $\del_\phi$ are intrinsic to the spheres $S_{s,\rho}$, in terms of $\del_s$, $\del_\rho$, $\Mb$, and $\overline{\Mb}$.

Since $\Mb$ is Lie dragged along the conformal geodesics while the frame (and hence $m$) is parallel with respect to the conformal connection, the vectors $\Mb$ and $m$ will in general differ at events past $\Sigma_0$. It is not even clear that $m$ will remain tangent to the spheres $S_{s,\rho}$. This discrepancy raises an issue:  we take components with respect to the space-time frame $(e_0,e_1,e_2,e_3)$ but we use the derivative operators defined with respect to the fiducial $\Mb$ and $\overline{\Mb}$. In order to consistently use the $\eth$ and $\eth'$ operator as defined above in terms of $\Mb$ and its complex conjugate we have to make sure that the components taken with the space-time frame nevertheless are properly spin-weighted when $\Mb$ is transformed according to~\eqref{eq:phasetrafo}. 

This is indeed the case as the following argument shows. We first note that both frame transports have the property that a frame rotated by an angle $\alpha$ is transported into a frame rotated by the same angle. Thus, a phase transformation of $\Mb$ by an angle $\alpha$ at an event $P_s$ with coordinates $(s,\rho,\theta,\phi)$ corresponds to a phase transformation of $\Mb$ by $\alpha$ at the point $P_0$ with coordinates $(0,\rho,\theta,\phi)$ on $\Sigma_0$, and hence to the same phase transformation of $m$ at $P_0$, which via parallel transport corresponds to a phase transformation by $\alpha$ for $m$ at $P_s$. In this way, the phase transformations of $\Mb$ and $m$ are tied together at every event and, therefore, the components taken with respect to space-time frame are properly spin-weighted under changes of $\Mb$, and we may apply the $\eth$-formalism to them. The same argument applies to the ``steered transport'' described in section~\ref{sec:conf-gauss-gauge} under certain circumstances that we will describe next.

In order to simplify the boundary treatment and the analysis on $\scri$ we steer the spatial part of the tetrad using the free infinitesimal rotation in such a way that it is adapted at every point in $M$ to the sphere through that point: we want $e_2$ and $e_3$ to remain tangent to the sphere. This implies that $e_2(\rho)=0=e_3(\rho)$ must be satisfied throughout the evolution. Taking a time derivative of $e_2(\rho)$ along the vector $u$ and using the frame transport equation~\eqref{eq:tetradtransportrotation} yields
\[
  \begin{aligned}
    \frac{\dd}{\dd s}(e_2(\rho)) &= u(e_2(\rho)) = [u,e_2](\rho) + e_2(u(\rho)) = \hnabla_u e_2(\rho) - \hnabla_{e_2}u(\rho) \\
    &= \omega^1{}_2e_1(\rho) + \omega^3{}_2e_3(\rho) - \hGamma_{02}{}^{1}e_1(\rho) - \hGamma_{02}{}^{2}e_2(\rho) - \hGamma_{02}{}^{3}e_3(\rho)
\end{aligned}
\]
and, similarly, for $e_3(\rho)$. If $e_2(\rho)=e_3(\rho)=0$ everywhere, then a part of the infinitesimal rotation is necessarily determined by
\begin{equation}
\omega^1{}_2=\hGamma_{02}{}^{1}, \quad \omega^1{}_3=\hGamma_{03}{}^{1}.\label{eq:14}
\end{equation}
On the other hand, if these conditions hold, then the equations derived above form a homogeneous linear system for $e_2(\rho)$ and $e_3(\rho)$ and hence these functions vanish everywhere if they vanish initially. Note, that these gauge conditions still do not fix the frame completely since $\omega^2{}_3$ remains undetermined, i.e., we are still free to arbitrarily rotate $e_2$ and $e_3$ in the tangent plane to the spheres. As a consequence the relationship between the complex vectors $m$ and $\Mb$ is fixed in exactly the same way as in the case without steering on the initial surface $\Sigma_0$ and the application of the $\eth$-formalism is consistent.

The conditions~\eqref{eq:14} have a small flow-on effect on the frame and tetrad equations in the GCFE system, but not on the tensorial equations for the Schouten and Weyl tensors.

At each point in $M$ we can expand the frame vectors $e_{c}$ in the coordinate basis $(\del_s,\del_\rho,\del_\theta,\del_\phi)$ or equivalently, in terms of $\Mb$ and $\overline{\Mb}$ in the form
\[
  \begin{aligned}
  e_{0} &= \del_s,\\
  e_{1} &= c_{1}^0\del_s + c_{1}^1\del_\rho + c_{1}\Mb + \bar{c}_{1}\overline{\Mb}  ,
\end{aligned}\qquad
\begin{aligned}
e_{2} &= c_{2}^0\del_s  + c_{2}^1\del_\rho  + c_{2}\Mb + \bar{c}_{2}\overline{\Mb}  ,\\
e_{3} &= c_{3}^0\del_s  + c_{3}^1\del_\rho  + c_{3}\Mb + \bar{c}_{3}\overline{\Mb}  ,\\
\end{aligned}
\]
When we assume the frame to be transported using steering, then the coefficients $c_2^1$ and $c_3^1$ vanish identically.

The preparation of the GCFE as given in (\ref{GCFEeq:1}--\ref{GCFEeq:4}) now proceeds in a straightforward way. Expand the equation with respect to the frame $(e_0,e_1,e_2,e_3)$, replace the directional derivatives along the frame vectors by the coordinate derivatives $\del_s$, $\del_\rho$ and the derivatives along $\Mb$ and finally replace the latter by the appropriate $\eth$ operators. This procedure yields a system of equations which is properly spin-weighted.

In our code we represent the spin-$s$ quantities $\eta$ as grid functions on the sphere sampled at points $(\theta_i,\phi_k)$ and we use 
\[
\Mb = \frac1{\sqrt2} \left( \del_\theta - \frac{\ii}{\sin\theta}\, \del_\phi \right).
\]
Then the connection coefficient $\alpha$ in~\eqref{eq:ethMb} becomes 
\[
\alpha = \frac{\cot\theta}{\sqrt2}.
\]
The code alternates between representing $\eta$ as a grid function or a truncated series~\eqref{eq:Ylmexp} with $s$ and $\rho$ dependent coefficients $\eta_{lm}$. The series representation is used to compute the derivatives along the sphere using~\eqref{eq:ethYlm}, while the grid representation is used to perform algebraic operations such as multiplication etc. These issues are explained in much more detail in~\cite{Beyer:2015ud,Beyer:2014bu,beyer2016numerical}.

\subsection{\label{subsec:spacespinorGCFE}The GCFE in the space-spinor formalism}

In this section we present the full GCFE in the space-spinor formalism~\cite{sommers1980space}. There are two reasons for this: first, the space-spinor formalism allows us to easily perform the time-space split of the equations into constraints and evolution equations by simply computing irreducible parts of the resulting spinor equations. Second, the formalism deals with complex quantities which means that the number of the actual equations after taking components is reduced compared to the tensorial equations by almost half. This leads to less numerical complexity and therefore ultimately also to lower round-off error (even though we have not checked that). Furthermore, it is easier to deal with symmetric trace-free quantities such as the electric and magnetic parts of the rescaled Weyl tensor because the trace-free condition is automatically satisfied in a spinor version compared to the tensorial version. Even though this means more algebraic work we feel that this is worthwhile. The space-spinor formalism is briefly summarized in the appendix~\ref{subsec:spacespinors}.

The idea is to express the covariant derivative operator $\hnabla_a$ in terms of $\nabla_a$, the Levi-Civita operator of the conformal metric $g_{ab}$ and the mediating 1-form $f_a$ (see~\eqref{eq:8}), then split them into evolution and constraint equations by decomposing the resulting equations. Once this is done the CGG is imposed and we obtain a complete system of evolution and constraint equations. We will not present the full derivations here (see~\cite{stevens2016numerical} for a full account), instead we give a brief summary.

The standard procedure is: %\footnote{Note that we have not yet fixed $t^a$ or the spin-frame, but this does not effect the procedure.}
\begin{itemize}
	\item Replace $\hnabla$'s with $\nabla$'s using the transformation laws~\eqref{eq:8} and \eqref{eq:9},
	\item convert the tensor fields to space-spinor fields
	\item convert $\nabla_a$ into $\del$ and $\del_{AB}$ as described in app.~\ref{subsec:spacespinors},
	\item decompose the equations into irreducible pieces to obtain evolution and constraint equations.
\end{itemize}
We apply this procedure to the field equations \eqref{GCFEeq:2} and \eqref{GCFEeq:3} for $\hP_{ab}$ and $K_{abc}{}^d$.  In doing so we will also change our notation slightly: the Schouten tensor $\hP_{ab}$ related to the Weyl connection is the only Schouten tensor appearing in the equations, so in order to avoid confusion with complex conjugation of space-spinors we will drop the hat. Thus, from now on $P_{ab}$ will denote the Schouten tensor for the Weyl connection.

To illustrate the procedure, we show how to treat the equation for $K_{abc}{}^d$. We first define $\psi_{ABCD}= \Theta^{-1}\Psi_{ABCD}$ as the only irreducible piece of $K_{abcd}$ in the spinor formalism. Skipping the first step as~\eqref{GCFEeq:3} is already in terms of the conformal connection, we convert tensor to spinor indices and then convert  primed to unprimed indices. Splitting the covariant derivative into $D$ and $D_{AB}$ derivatives and expressing these in terms of $\del$ and $\del_{AB}$ yields the equation
\begin{equation*}
  -\frac12 \del \psi_{ABCD} - K_{(A}{}^E \psi_{BCD)E} + \del_D{}^E \psi_{ABCE} + \frac32 K_D{}^E{}_{(A}{}^F \psi_{BC)EF} + \frac12 K_D{}^E{}_{E}{}^F\psi_{ABCF} = 0.
\end{equation*}
This equation is symmetric in $ABC$. Taking the totally symmetric part gives an evolution equation for $\psi_{ABCD}$
\begin{equation*}
  \del\psi_{ABCD} - 2\del_{(A}{}^E\psi_{BCD)E} = -2K_{(A}{}^E\psi_{BCD)E} + 3K_{(A}{}^E{}_B{}^F\psi_{CD)EF} - K_{E(A}{}^{EF}\psi_{BCD)F},
\end{equation*}
while the anti-symmetric part gives us a constraint
\begin{equation*}
  \del^{CD}\psi_{ABCD} = -K^{CE}{}_E{}^D\psi_{ABCD} - K^{CDE}{}_{(A}\psi_{B)CDE}.
\end{equation*}

Up to this point the choice of $t^a$ was irrelevant. However, to derive the gauge-related equations from the remaining equations \eqref{GCFEeq:1} and \eqref{GCFEeq:2}, we must fix the time-like vector $t^a$ and derive the spinor form of the CGG equations. We fix $t^a$ to be proportional to the tangent vector $u^a$ of the curves. Then the spin-frame $(o_A,\iota_A)$ corresponding to the tetrad must satisfy the usual normalization condition $o_A\iota^A=1$ and we impose the condition $t_{AA'} = o_Ao_{A'} + \iota_A\iota_{A'}$. Then we have
\[
\hat o_A = \iota_A, \qquad \hat\iota_{A} = - o_A,
\]
and the transport equations~\eqref{eq:tetradtransport} or \eqref{eq:tetradtransportrotation} for the frame imply equations for the spin frame. We will explicitly discuss here the equations for the pure CGG without frame rotations, the other case being very similar. Thus, the spin frame must satisfy  the equation $t^a\hnabla_ao_B =0$ (and its complex conjugate). We define the components of the frame vectors in terms of the coordinate basis by
\begin{equation*}
c^\mu:=\del x^\mu, \qquad c_{AB}^\mu := \del_{AB} x^\mu.
\end{equation*}
Since both $t^a$ and the time-like frame vector are parallel to $u^a$ we have
\[
u^a = e^a_\c{0} =\frac{1}{\sqrt{2}}t^a.
\]
The gauge conditions then imply
\begin{equation*}
c^\mu = \sqrt{2}\delta_0^\mu.
\end{equation*}
Furthermore, we define the spinor fields
\begin{equation*}
\gamma_A:=\del o_A  \implies \hat\gamma_A = \del \iota_A, \qquad 
\gamma_{ABC}:=\del_{AB} o_C   \implies \hat\gamma_{ABC} = -\del_{AB} \iota_C .
\end{equation*}
They satisfy the relations
\[
\gamma_C \iota^C = \hat\gamma_C o^C,\qquad 
\gamma_{ABC} \iota^C = -\hat\gamma_{ABC} o^C.
\]
These spinor fields encode part of the connection coefficients $\widehat\Gamma_{\c{ab}}{}^{\c{c}}$, the other pieces being contained in $K_{AB}$, $K_{ABCD}$ and~$f_{AB}$.

The tetrad is parallel along the conformal geodesics so that we have $u^a\hnabla_a g_{bc}=0$ from which we conclude together with eq.~\eqref{eq:1} that
\[
t^a\nabla_a g_{bc} = -2t^{AA'}f_{AA'} g_{bc} = 0 \implies f_{AB}=f_{BA}.
\]
Furthermore, 
\[
0 = t^a \hnabla_a o_C = t^a\nabla_a o_C - t^{AA'} f_{CA'} o_A = \del o_C + \frac12 K_C{}^B o_B - f_C{}^A o_A .
\]
This and the analogous equation for $\iota_C$ yield the equations
\begin{equation*}
\begin{aligned}
  \gamma_C + \frac12\left( K_C{}^B - 2 f_C{}^B\right)o_B  = 0,\\
  \hat\gamma_C + \frac12\left( K_C{}^B - 2 f_C{}^B\right)\iota_B = 0,
\end{aligned}
\end{equation*}
which imply
\begin{equation}
 K_{AB} = 2f_{AB}, \qquad \gamma_C = 0. 
\end{equation}
The last consequence of the gauge conditions involves the Schouten spinor of the Weyl connection, which we denote in the space-spinor formalism as $P_{ABCD}:= t_B{}^{A'}t_D{}^{B'} P_{AA'CB'}$ (again note, that we dropped the hat here to avoid confusion with complex conjugation as mentioned above). This spinor field satisfies the reality condition
\begin{equation*}
	\widehat{P}_{ABCD} = P_{BADC},
\end{equation*}
and the gauge condition~\eqref{eq:uSchouten} implies
\begin{equation*}
	t^{AA'}P_{AA'BB'} = 0 \implies P_{ABCD} = P_{BACD}.
\end{equation*}
This concludes the translation of the CGG to the space spinor formalism. We can now proceed to translate the tensorial equations to spinorial form following the procedure outlined above. Incorporating the gauge conditions we arrive at a fully determined system for the unknowns
\begin{equation}
(c^\mu_{AB}, \gamma_{ABC}, K_{ABCD}, f_{AB}, P_{ABCD}, \psi_{ABCD}) \label{unknowns}
\end{equation}
in which $\Theta$ and the components of $h_{AB}$ are considered known since they are determined \emph{a priori}  from initial data by the  equations~\eqref{eq:10} and~\eqref{eq:5}.

The evolution equations are
\begin{subequations}\label{eveqs}
\begin{align}
  \del c^0_{AB} &= -\sqrt{2}f_{AB} - K_{AB}{}^{CD}c^0_{CD}, \label{ev:6} \\
  \del c^i_{AB} &= -K_{AB}{}^{CD}c^i_{CD}, \qquad i=1,2,3. \label{ev:7} \\
  \del K_{ABCD} &= -K_{AB}{}^{EF}K_{EFCD}  - 2P_{AB(CD)} + \Theta\psi_{ABCD} + \Theta\hat\psi_{ABCD}, \label{ev:3} \\
  \del\gamma_{ABC} &= -K_{AB}{}^{EF}\gamma_{EFC} - o_{(A}K_{B)CDE}f^{DE} + \frac12o_CK_{ABEF}f^{EF} +
  K_{ABE(C}f_{D)}{}^Eo^D \nonumber \\
  &\hskip6em + \frac12P_{ABE}{}^Eo_C + \eps_{C(A}P_{B)DE}{}^Eo^D - \frac12\Theta\psi_{ABCD}o^D + \frac12\Theta\hat\psi_{ABCD}o^D , \label{ev:5} \\
  \del f_{AB} &= -K_{ABEF}f^{EF} + P_{ABC}{}^C, \label{ev:4} \\
  \del P_{ABCD} &= -K_{ABEF}P^{EF}{}_{CD} + \psi_{ABCE}h^E{}_D - \hat\psi_{ABDE}h_C{}^E, \label{ev:1} \\
  \del\psi_{ABCD} &= 2\del_{(A}{}^E\psi_{BCD)E} - 2K_{(A}{}^E\psi_{BCD)E} + 3K_{(A}{}^E{}_B{}^F\psi_{CD)EF} - K_{E(A}{}^{EF}\psi_{BCD)F}, \label{ev:2} 
\end{align}
\end{subequations}
The constraint equations express the vanishing of the following ``zero quantities''
\begin{subequations}\label{constreqs}
\begin{align}
	0 &= Z^0_{AB} := \del^C{}_{(A}c^0_{B)C} + \frac{1}{\sqrt{2}}K_{(A}{}^C{}_{B)C},\label{constr:6}  \\
	0 &= Z^i_{AB} := \del^C{}_{(A}c^i_{B)C}, \qquad i=1,2,3.\label{constr:7}\\
	0 &= J_{ABC}  := -\del_{(A}{}^E\gamma_{B)EC} + \frac14f_{AB}f_{CD}o^D + \frac12K_{(A}{}^E{}_{|C|}{}^DK_{B)EFD}o^F - 
		\frac18\left( P_{AC(BD)} + P_{BC(AD)} \right)o^D \nonumber \\
	&\hskip4em 	- \frac18\left( P_{AD(BC)} + P_{BD(AC)} \right)o^D + \frac14\left( P_A{}^D{}_{BD} + P_B{}^D{}_{AD} \right)o_C 
		+ \frac18\left( P_A{}^D{}_{(CD)}o_B + P_B{}^D{}_{(CD)}o_A \right) \nonumber \\
	&\hskip4em 	- \frac18\left( \eps_{AC}P_B{}^E{}_{(DE)} + \eps_{BC}P_A{}^E{}_{(DE)} \right)o^D - 
		\frac12o^D\del_{D(A}f_{B)C} + \frac12o_{(A}\del_{B)}{}^Ef_{CE},\label{constr:5}  \\
	0 &= Z_{ABCD} := -\del_{(A}{}^EK_{B)ECD} - \frac12f_{CD}K_{(A}{}^E{}_{B)E} - \frac12f_{C(A}K_{B)}{}^E{}_{DE} - 
	  	\frac12\eps_{C(A}f^{EF}K_{B)DEF} \nonumber \\
	  &\hskip4em 	- \frac12\eps_{D(A}f^{EF}K_{B)ECF} + \frac12\eps_{C(A}P_{B)DE}{}^E + \frac12\eps_{D(A}P_{B)CE}{}^E + 
	  	\frac12\Theta\psi_{ABCD} - \frac12\Theta\hat\psi_{ABCD},\label{constr:3} \\
	0 &= T_{AB}   := \del_{(A}{}^Ef_{B)E} + \frac12P_{(A}{}^E{}_{B)E} - \frac12P_{E(A}{}^E{}_{B)}, \label{constr:4}\\
	0 &= U_{ABCD} := -\del_{(A}{}^EP_{B)ECD} - \frac12f_{CE}P_{(A}{}^E{}_{B)D} - \frac12f_{DF}P_{(A}{}^E{}_{|C|B)} - 
	    \frac12\left( f_{D(A}P_{B)}{}^E{}_{CE} + f_{C(A}P_{B)}{}^E{}_{ED} \right) \nonumber \\
	    &\hskip4em + \frac12P_{(A}{}^E{}_{|C|}{}^FK_{B)EDF} - \frac12P_{(A}{}^{EF}{}_{|D|}K_{B)ECF} + 
	    \frac12\psi_{ABCE}h^E{}_D + \frac12\hat\psi_{ABDE}h_C{}^E, \label{constr:1}\\
	0 &= G_{AB}   := \del^{CD}\psi_{ABCD} + K^{CE}{}_E{}^D\psi_{ABCD} + K^{CDE}{}_{(A}\psi_{B)CDE} \label{constr:2}.
\end{align}
\end{subequations}

In order to impose the $\eth$-calculus on our system, we replace the expansion $\del_{AB}=c^\mu_{AB}\del_\mu$ with
\begin{equation*}
	\del_{AB} = c^0_{AB}\del_s + c^1_{AB}\del_\rho - 
	\Big{(}\frac1R\iota_A\iota_B + 2Xo_{(A}\iota_{B)} + Yo_Ao_B\Big{)}\Mb + 
	\Big{(}\frac1Ro_Ao_B - 2\bar Xo_{(A}\iota_{B)} + \bar Y\iota_A\iota_B\Big{)}\overline{\Mb},
\end{equation*}
where $X$ and $Y$ are complex functions while $R$ can be chosen to be real as a consequence of the remaining gauge freedom on the spheres. This is essentially a repackaging of the quantities $c^2_{AB}$ and $c^3_{AB}$. It is now a straightforward but tedious procedure to replace the derivatives along $\Mb$ and its complex conjugate in terms of $\eth$ and $\eth'$ using~\eqref{eq:ethMb}, which ultimately yields properly spin-weighted equations for the spinor components of the fields. The equations for the new functions $R$, $X$ and $Y$ come as evolution equations
\begin{subequations}\label{RXYevs}
	\begin{align}
		\del_t R &= \frac{1}{\sqrt{2}}R K_{02} + \sqrt{2} R^2 X K_{01} + \frac{1}{\sqrt{2}}R^2YK_{00},\label{Rev} \\
		\del_t X &= \frac{1}{\sqrt{2} R} K_{12} + \sqrt{2} X K_{11} + \frac{1}{\sqrt{2}}YK_{10},\label{Xev} \\
		\del_t Y &= -\frac{1}{\sqrt{2}R}K_{22} - \sqrt{2} R X K_{21} - \frac{1}{\sqrt{2}}YK_{20},\label{Yev}
	\end{align}
\end{subequations}
and constraint equations
\begin{subequations}\label{newconstrs}
	\begin{gather}
		K_{02} - K_{20} + 2 R \Big{(}XK_{01} + \bar X K_{21}\Big{)} + R\Big{(}YK_{00} - \bar YK_{22}\Big{)} = 0, \label{constrnew1} \\[12pt]
		\Big{(}c^1{}_2 + \frac12 RYc^1{}_0\Big{)}\del_rX + \frac12RYc^1{}_2\del_r\bar{X} + c^1{}_1\del_rY
		- \frac{3}{2\sqrt{2}}Y\eth X - \frac{1}{2\sqrt{2}}RY^2\eth\bar{X} + \frac{1}{\sqrt{2}}X\eth Y \nonumber \\
		+ \Big{(}\frac{1}{\sqrt{2R}} + \frac{1}{2\sqrt{2}}RY\bar{Y}\Big{)}\eth'X + \frac{1}{2\sqrt{2}}Y\eth'\bar{X} + 
		\frac{1}{\sqrt{2}}\bar{X}\eth'Y + \text{algebraic terms} = 0, \label{constrnew2}\\[12pt]
		\frac{2}{R}c^1{}_1\del_rR - R c^1{}_0R\del_rX + Rc^1{}_2\del_r\bar{X} 
		+ \frac{\sqrt{2}}{R}X\eth R + \frac{1}{\sqrt{2}}\eth X - \frac{1}{\sqrt{2}}RY\eth\bar{X}
		+ \frac{\sqrt{2}}{R}\bar{X}\eth'R \nonumber \\
		- \frac{1}{\sqrt{2}}R\bar{Y}\eth'X + \frac{1}{\sqrt{2}}\eth'\bar{X} + \text{algebraic terms} = 0,\label{constrnew3}
	\end{gather}
\end{subequations}
where the ``algebraic terms'' are quite lengthy containing components of $K_{ABCD}$, the spin-weighted components of $\gamma_{ABC}$ and the frame components.

The new proper spin-weighted system then comprises of the evolution equations, constraint equations and explicit expressions for $\Theta$ and $h_a$ given in \eqref{eveqs}, \eqref{constreqs} \eqref{eq:10} and \eqref{eq:5} but now the equations for $c^2_{AB}$ and $c^3_{AB}$ are replaced by equations for $R,\;X$ and $Y$ given in \eqref{RXYevs} and \eqref{newconstrs}.

As a partial justification for the correctness of the final system, we expressed the Minkowski and Anti-de Sitter space-times  analytically in the CGG adapted to spherical symmetry. The unknowns of the system were then computed from these solutions and were shown to satisfy both the evolution and constraint equations identically.

\subsection{\label{subsec:subsidiarysystem}The subsidiary system}

The constraint equations derived from the GCFE propagate on an analytical level, as shown by Friedrich~\cite{friedrich1995einstein}, in the sense that the ``zero-quantities'' themselves satisfy a semi-linear homogeneous symmetric hyperbolic system of PDE. For the initial value problem this implies that if these quantities vanish initially then they vanish everywhere due to the uniqueness of solutions. Constraint propagation at the numerical level is another story entirely. Clearly, the violation of the constraints, i.e., non-vanishing ``zero-quantities'' are governed by this propagation system. Therefore, it is important to understand the properties of this system since it has implications on the stability and hence on the well-posedness of the evolution. 

We have derived the subsidiary system for all the constraints in our formulation. This is a straightforward but awkward process that we will not elaborate here. Instead, we list only the propagation equation for the constraints arising from the system for the rescaled Weyl spinor since this is the only one needed in the sequel. 
\begin{equation}
\begin{aligned}
  \del G_{AB} = \del_{(A}{}^CG_{B)C} &- \frac32K_{(A}{}^CG_{B)C} + \frac12\tK_{(A}{}^CG_{B)C} \\
  &+ \frac12G^{CD}\tK_{ABCD} - \frac43\tK G_{AB} - T^{CD}\psi_{ABCD} - \frac12Z_{(A}{}^{CDE}\psi_{B)CDE} - 2Z^{ECD}{}_E\psi_{ABCD}. \label{eq:12}
\end{aligned}
\end{equation}
Here, $\tK_{ABCD}$, $\tK_{AB}$ and $\tK$ are the irreducible parts of $K_{ABCD}$. The complete subsidiary system can be found in appendix~\ref{sec:compl-subs-syst}.

\section{\label{sec:ibvp}The IBVP framework}

We now suppose, and will assume for the rest of the document, that the CGG has the coordinate system $\{s,\rho,\theta,\phi\}$. This is to make the distinction between the coordinates used in the CGG gauge and other frequently used coordinates such as $t$ and $r$, which we reserve for exact solutions such as the Schwarzschild space-time.

\subsection{\label{subsec:maxbc}Maximally dissipative boundary conditions}

We choose the spatial extent of the computational domain to be defined by $\rho \in [\rho_0, \rho_1] =: I$ resulting in the cylindrical domain $I \times S^2$. Due to our choice of coordinates, the boundary of the domain is given by the equations $\rho=\rho_0$ on the left boundary and $\rho=\rho_1$ on the right boundary, defining two hyper-surfaces which (at least for some time) are time-like. Inspection of the evolution equations \eqref{eveqs} and \eqref{RXYevs} shows that the only equation which needs boundary conditions is the evolution equation \eqref{ev:2} for $\psi_{ABCD}$. All other equations propagate along the boundary.

It is well known that the field equation for the gravitational spinor admits a symmetric hyperbolic system of PDEs (see for example \cite{friedrich2004smoothness}, \cite{friedrich1996hyperbolic}). The equation \eqref{ev:2} is already in symmetric hyperbolic form. This is best verified by checking that the symbol of the equation is Hermitian. For any pair $(p,p_{AB})$ of a positive real number $p$ and real symmetric spinor $p_{AB}$ consider the sesquilinear form on totally symmetric spinors of rank 4 defined by 
\[
  \left<\chi,\psi\right> = \widehat\chi^{ABCD}p \psi_{ABCD} - 2 \widehat\chi^{ABCD}p_A{}^E\psi_{BCDE}.
\]
It is obtained by replacing the derivative operators in the principal part of \eqref{ev:2} by $p$ and $p_{AB}$, respectively. Now it is easy to see that this form is Hermitian, i.e., that
\[
\overline{\left<\chi,\psi\right>} = \left<\psi,\chi\right>
\]
for any choice of $(p,p_{AB})$ and positive definite for $p_{AB}=0$. Therefore, the system  \eqref{ev:2} is symmetric hyperbolic.

To impose boundary conditions it is useful to first analyse the characteristics of the system near the boundaries. We choose the right boundary here, the left being treated analogously. We are interested in the characteristics that cross the time-like hyper-surface defined by $\rho=\rho_2$. Since the system is symmetric hyperbolic, at each point $P$ of the boundary there is a family of characteristic cones with vertex $P$ opening towards the past along which the field propagates with characteristic speeds. We want to find the intersection of these cones with the 2-dimensional plane spanned by $\del_s$ and $\del_\rho$. These will be lines with slopes given by the characteristic speeds $\lambda$. They are obtained from a generalized eigen-value problem which arises from the principal part of \eqref{ev:2} by inserting the ansatz $\psi_{ABCD} = X_{ABCD}\, \ee^{\ii(\rho-\lambda s)}$. Then the characteristic speeds $\lambda$ are obtained as those values for which the algebraic system 
\[
\left[\frac{\lambda}{\sqrt2}\delta_{(A}{}^E - \lambda c^0_{(A}{}^E + c^1_{(A}{}^E \right] X_{BCD)E} = 0
\]
has non-trivial solutions. Define the spinor $T_{AB} = \frac12T \eps_{AB} + S_{AB}$ with $T=\sqrt2\lambda$ and $S_{AB} =   c^1_{AB} - \lambda c^0_{AB}$. Near the initial hyper-surface we can write $S_{AB} = s \alpha_{(A} \beta_{B)}$ in terms of its principal null spinors with $\alpha_A\beta^A =1$. Then we find that the system has non-trivial solutions only if
\[
2T + (2-k) s = 0, \qquad \text{with } \quad k=0,\ldots, 4
\]
and we find the resulting speeds
\begin{equation}
\lambda_{0,4} = 
\frac{- f^{01} + \mathrm{sign}(k-2) \sqrt{(f^{01})^2 - f^{11}(1+f^{00})}}{1+f^{00}},\quad
\lambda_{1,3}=\frac{- f^{01} + \mathrm{sign}(k-2) \sqrt{(f^{01})^2 - f^{11}(4+f^{00})}}{4+f^{00}},\quad
\lambda_2 = 0.\label{psi_characteristics}
\end{equation}
Here, we have defined $f^{ij}=c_{AB}^i c^{jAB}$ where $f^{ii}<0$. The corresponding solutions $X_{ABCD}^k$ are simply the totally symmetric outer products of the principal spinors $\alpha_A$ and $\beta_A$ with $k$ giving the number of $\beta_A$ in the product. Thus, $X_{ABCD}^1 = \alpha_{(A}\alpha_{B}\alpha_{C}\beta_{D)}$ etc. It is not difficult to see that the lines defined by $\lambda_0$ and $\lambda_4$ correspond to the intersection of the light-cone with the plane spanned by $\del_s$ and $\del_\rho$, while $\lambda_1$ and $\lambda_3$ correspond to a time-like cone inside the light-cone. Thus, the components $X^0_{ABCD}$ and $X^4_{ABCD}$ propagate with the speed of light.

We decompose the spinor $\psi_{ABCD}$ with respect to the spin-frame $(\alpha_A,\beta_A)$, writing $\psi_{ABCD} = \sum_{k=0}^4 \tpsi_k X^k_{ABCD}$.
Since $\lambda_k < 0$ for $k=0,1$, the corresponding components $\tpsi_k$ propagate towards decreasing $\rho$ while $\tpsi_3$ and $\tpsi_4$ travel towards increasing $\rho$. The component $\tpsi_2$ propagates along the boundary. Thus, on the right boundary we need to provide boundary conditions for $\tpsi_0$ and $\tpsi_4$. 

A very general class of boundary conditions are the maximally dissipative boundary conditions, which express the incoming components in terms of the outgoing ones and freely specifiable functions on the boundary. In our case on the right boundary we write
\begin{equation}\label{eq:33}
 	\begin{pmatrix}
  		\widetilde{\psi}_0 \\
 		\widetilde{\psi}_1 
 	\end{pmatrix} = 
	\begin{pmatrix}
  		q_0 \\
 		q_1 
 	\end{pmatrix}
        + H
 	\begin{pmatrix}
  		\widetilde{\psi}_3 \\
 		\widetilde{\psi}_4 
 	\end{pmatrix}.
\end{equation}
where the $q_i$ are free boundary data and $H$ is a $2\times2$ matrix defined on the boundary satisfying 
\begin{equation}\label{eq:32}
	H^* H \leq I_2,
\end{equation}
where $I_2$ is the identity matrix. The matrix $H$ encodes the reflective properties of the boundary and the inequality ensures that the reflected `energy' does not exceed the `energy' hitting the boundary. We will use the simplest case $H=0$, which corresponds to the boundary being completely transparent to the outgoing modes. 

In order to implement these conditions we need to run through the following procedure at every point on the boundary: determine the principal null directions of the spinor $S_{AB}$, compute the characteristic speeds, project out the outgoing characteristic modes from $\psi_{ABCD}$, use the boundary condition to compute the ingoing characteristic modes, and then reassemble the spinor $\psi_{ABCD}$ on the boundary. This yields a system of the form
\begin{equation}
  \label{eq:15}
  q_k = \sum_{l=0}^4 M^l{}_k \psi_l ,\qquad k=0,1.
\end{equation}
Here, $M$ is a $2\times5$-matrix made up from the components of the spin-frame transformation between $(o_A,\iota_A)$ and $(\alpha_A,\beta_A)$. Then, imposing the boundary conditions means solving this system for $\psi_0$ and $\psi_1$, regarding the other components as given with free functions $q_k$.

Note, that the expressions for $\lambda_0$ and $\lambda_4$ become singular when $1+f^{00}=0$. Since $g(\dd s,\dd s) = g^{ss} = 1+f^{00}$ this will happen, when the hyper-surface $s=\mathrm{const}$ becomes null at the boundary. Due to our choice of gauge we cannot prevent this from happening. However when the frame components $c^0_{AB}$ vanish for all time, as discussed in section \ref{subsec:MinkowskiIBVP}, we are in a similar situation to the Friedrich-Nagy gauge \cite{friedrich1999initial} where the spatial frame vector normal to the boundary remains normal for all time. We will have more to say about these issues later.

The characteristics for this particular choice of evolution system are simple, there will only ever be two ingoing modes to provide boundary conditions for. However in general, this will not be the case. Adding combinations of the constraints \eqref{constreqs} to the evolution equations will alter their characteristics, and hence the number of ingoing modes. The characteristic speeds are functions of the frame components, and hence the number of ingoing modes may change over time.

\subsection{\label{sec:constrpreservingbc}Constraint preserving boundary conditions}

In the previous section we employed maximally dissipative boundary conditions to obtain a stable evolution system, however this did not take into account the propagation of the constraint system. As the constraint equations coming from the field equation for $\psi_{ABCD}$ are PDEs, if they are violated on the boundary these violations may propagate into the interior. As one can see in the subsidiary system given in section \ref{subsec:subsidiarysystem}, many of the constraint propagation equations in the subsidiary system are written in terms of the gravitational spinor's constraints. Thus if they are violated, they will cause other constraints to be violated too.

At each boundary, the maximal dissipative boundary formalism yields two degrees of freedom for the boundary data for our evolution system. We, however, expect only one physical degree of freedom. This is most likely the ingoing mode travelling on the light-cone, i.e., $\tpsi_0$ or $\tpsi_4$ depending on the boundary. The discrepancy in the number of degrees of freedom is related to the violation of the constraints and we will use the additional ingoing mode to kill ingoing constraint violating modes. In order to define the required constraint preserving boundary conditions for our evolution system, we need to analyse the subsidiary system.

The principal part of the subsidiary equation for the gravitational spinor constraint $G_{AB}$ \eqref{subsideq3} is
\begin{equation}\label{eq:6}
	\del G_{AB} = \del^C{}_{(A}G_{B)C}.
\end{equation}
This system is again symmetric hyperbolic and we can go through the same analysis as we did on the $\psi_{ABCD}$ evolution system in the previous section. We find that there are three characteristic modes propagating with three different characteristic speeds. The remarkable fact is that these three speeds agree with the characteristic speeds $\lambda_1$, $\lambda_2$ and $\lambda_3$ of the $\psi_{ABCD}$ system, see~\eqref{psi_characteristics}. Furthermore, the characteristic modes are again given in terms of the principal null spinors $\alpha_A$ and $\beta_A$, being proportional to $\alpha_A\alpha_B$, $\alpha_{(A}\beta_{B)}$, and $\beta_A\beta_B$, respectively. Thus, the constraints propagate on characteristics which agree with the time-like characteristics of the Weyl system. In order to kill the ingoing constraint violating mode on the right boundary we need to ensure that the ingoing component proportional to $\alpha_A\alpha_B$ vanishes on the boundary.

The problem is then to find a way to prescribe boundary data for the $\psi_{ABCD}$ system on the boundary which gives interesting behaviour as well as having no ingoing mode in the subsidiary system. Since the physically relevant quantities at the boundary are  $\psi_0$ resp. $\psi_4$ one may ask whether it is possible to leave their prescription free, while fixing the remaining freedom in choosing $\psi_1$ resp. $\psi_3$ in such a way that the ingoing mode of the subsidiary system is zero. This is in fact possible, and is the main result of the paper.

Our proposed resolution on the right boundary (the left is analogous) for the case of $\psi_0,\psi_1$ ingoing, $\psi_3,\psi_4$ outgoing and $\psi_2$ propagating along the boundary is to derive an ODE for the free datum $q_1$ as follows:
\begin{itemize}
	\item Regard $\psi_0$ as a free function, i.e., the boundary data $q_0$ is a free spin-2 function.
	\item Take a time derivative of the boundary equations \eqref{eq:15} for $\psi_0$ and $\psi_1$ and replace all the time derivatives with their evolution equations. This will result in two equations containing the terms $\del_s q_1, \del_\rho \psi_0$ and $\del_\rho \psi_1$.
	\item Solve these equations simultaneously for $\del_\rho \psi_0$ and $\del_\rho \psi_1$.
	\item Now consider the single equation which requires the vanishing of the ingoing constraint violating mode. This equation also involves $\del_\rho \psi_0$ and $\del_\rho \psi_1$. Replace them with the expressions from the previous step.
	\item Finally, solve this equation for $\del_s q_1$.
\end{itemize}
The resulting equation is free of $\del_\rho\psi_0$ and $\del_\rho\psi_1$ and gives us an ODE for $q_1$ on the boundary. Thus, evolving this equation along the boundary yields $q_2$ which is used as the additional ``unphysical'' degree of freedom for the evolution  that should kill the ingoing mode from the subsidiary system.

The final remark here is how to extend this approach to incorporate the changing propagation directions of the $\psi_{ABCD}$ components if another evolution system was chosen with different characteristics. We note that there is a relationship between the characteristics of the evolution system and the subsidiary system, see for example \cite{gundlach2004symmetric}. This relationship is that there are always $n-1$ ingoing modes for the subsidiary system and $n$ ingoing modes for the evolution system. Hence there is always a degree of freedom to choose ingoing gravitational radiation for example, while just the right number of ingoing modes remaining to kill those of the subsidiary system.

At this point it might be useful to take count of the degrees of freedom in the system. As in every covariant geometric system of equations we have the usual gauge freedom of the choice of coordinates and basis of the tangent vectors. Since we formulate our equations in a conformally covariant way we have further gauge freedom, namely the choice of a conformal factor and of the conformal Weyl connection. All these choices have to be fixed in order to get a well-posed evolution system. In our approach they are all satisfied by imposing the Gauss gauge and choosing appropriate initial data for coordinates, frame, conformal factor and Weyl 1-form. Given our choice of evolution system, this reduces the freedom to two complex functions on each boundary, describing the incoming modes. However, as discussed above, one of these functions is determined by the requirement that the ingoing constraint violating mode should vanish. This leaves us with exactly one free function on each boundary, the ingoing spin-2 component of the gravitational field. This is a complex-valued function of spin-weight $\pm2$, so it encodes exactly the two polarization degrees of freedom in a gravitational wave. Note that when a different evolution system is chosen, the relationship between the evolution and subsidiary systems described in the previous paragraph ensures that there is still only ever one complex valued degree of freedom on each boundary.

\section{\label{sec:numerics}Numerics}

The aim of this section is to formulate a numerical implementation of the IBVP for the GCFE system and apply it in a variety of situations. The idea is to choose an initial space-like hyper-surface and prescribe on it data for the unknowns~\eqref{unknowns} so that the constraints \eqref{constreqs} are satisfied. Then we evolve the initial data using the evolution equations \eqref{eveqs} using boundary conditions (where applicable) according to process proposed in the previous section so the constraints remain satisfied during the evolution.

We must first discretize the continuous version of the equations in order to evolve the system on the computer using various numerical methods. We will discretize our system using the \textit{method of lines}. In full generality, we need to discretize three spatial dimensions, two of which are tangent to unit 2-spheres \footnote{This is because we introduced the unit-sphere derivative operators~$\eth$, $\eth'$.} while the third dimension is transverse to them. The action of the spherical operators $\eth$, $\eth'$ can be approximated using \textit{pseudo-spectral methods} using the algorithm given in~\cite{beyer2016numerical}. The radial direction is approximated using a straightforward fourth-order \textit{finite difference} method with the summation-by-parts property as described in \cite{strand1994summation}. Discretizing in this way yields a  semi-discrete system of ODEs in time, which we solve using a standard \textit{fourth-order explicit Runge-Kutta} scheme. In order to impose stable boundary conditions, we utilize the \textit{simultaneous-approximation-term} (SAT) method that is described in~\cite{carpenter1999stable}.

Before carrying out a numerical investigation into the validity of this scheme, we first tested the system and numerical methods by solving several IVP's for the Kottler space-time. This was done using different values and signs of the cosmological constant and by using the setup detailed in section~\ref{subsec:schwarzschildIBVP}. As these space-times are spherically symmetric, we found that our evolution and subsidiary systems reduced to ODEs in time, hence boundary conditions were not required. We discovered that we could reproduce the conformal structure of Schwarzschild, Schwarzschild-de Sitter (for the case of distinct cosmological and event horizons) and Schwarzschild-Anti-de Sitter space-times and the constraints were shown to propagate.

\begin{figure*}[htp]
  \centering 
  \includegraphics[scale=0.38]{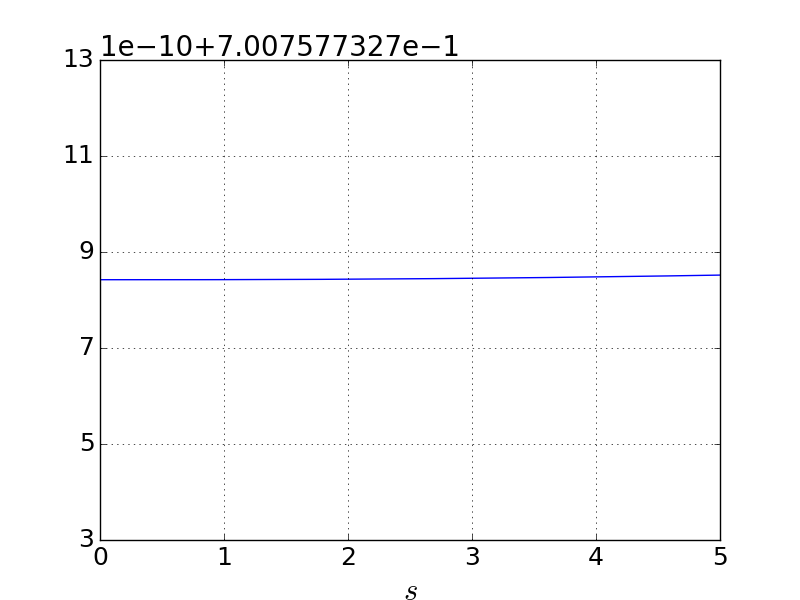}
  \caption{\label{fig:noisetest}A single noise test using randomly perturbed Minkowski initial and boundary data, where the curve represents the ``global'' $l_2$-norm of the system variables over time.}
\end{figure*}

We also conducted noise tests for the evolution system to test for numerical stability by checking that high frequency perturbations did not trigger non-linear growth. To this end we followed the apples with apples robustness test \cite{alcubierre2003towards} by using randomly perturbed Minkowskian initial and boundary data. We ran multiple simulations in axi-symmetry using $\rho$ and $\theta$ resolutions of $200$ and $32$ respectively with a timestep of $0.0025$ for $2000$ iterations. We calculated a ``global'' $l_2$-norm on each timeslice by summing the squares of each variable on each point, dividing by the number of points and then taking the square root. As can be seen in FIG.~\ref{fig:noisetest} this did not grow substantially over time, and for all practical purposes is negligible. This gives a good indication of numerical stability.

\subsection{\label{subsec:MinkowskiIBVP}The IBVP for non-linear gravitational perturbations of Minkowski space-time}

The next test concerns the boundary implementation. In particular, we need to check to what extent the method we discussed in section~\ref{sec:constrpreservingbc} actually kills ingoing constraint violating modes. We choose Minkowski initial data and imposing appropriate boundary conditions to shoot in gravitational waves, represented by $\psi_0$ and $\psi_4$ (propagating in the $\rho$-direction from right to left and left to right respectively). The boundary conditions are chosen to be axi-symmetric, i.e.\ independent of~$\phi$. The conformal Gauss gauge preserves the $\phi$-independence throughout the evolution. Hence we have a $2+1$ problem.

We will consider the physical representation of a part of Minkowski space-time with metric
\[
  g = \dd t^2 - \dd \rho^2 - \rho^2\left( \dd\theta^2 + \sin^2{\theta}\,\dd\phi^2\right)
\]
and $\rho \in [\rho_1,\rho_2]$ for some values $\rho_1<\rho_2$ and then perturb it with incoming gravitational waves. To choose the gauge conditions we take the conformal factor and the 1-form $h_a$ from the exact Minkowski space-time and impose them also in the perturbed space-time. Since we look at the physical representation of the Minkowski metric we have $\Theta=1$ and $h_a=0$. These are also the initial conditions for $\Theta$ and $h_a$. 
This choice says that the gauge is adapted to metric time-like geodesics and it has the consequence the frame components $c^0_{AB}$ vanish for all time, which can be seen from the evolution equations. Therefore, there is no $\del_s$ contribution from the expansion of $\del_{AB}$ during the evolution, so that the spatial frame vectors will stay tangent to the $s=\text{constant}$ hyper-surfaces.
%We must stress here that the condition $c^0_{AB}=0$ holds for the entire evolution can not in general be satisfied, it is only with an appropriate choice of gauge and initial data that this works here. Hence our boundary treatment must in general vear away from the Friedrich-Nagy situation, i.e.\ when the frame orthogonal to the boundary remains orthogonal.

The initial metric is
\[
	h = -\text{d}\rho^2 - \rho^2\Big{(}\text{d}\theta^2 + \sin^2{\theta}\text{d}\phi^2\Big{)},
\]
and with the choice of initial triad as indicated in section~\ref{subsec:ethoperators} we find the only non-vanishing system components to be
\[
	R = \rho,\qquad c^1{}_1 = \frac{1}{\sqrt{2}},\qquad \gamma_{20} = \hat\gamma_{01} = -\frac{1}{\sqrt{2}\;\rho}.
\]
The characteristic speeds of the components of $\psi_{ABCD}$ used in the evolution are given by
\begin{equation*}
	-\sqrt{2}c^1{}_1,\qquad	-\frac{c^1{}_1}{\sqrt{2}}, \qquad 0, \qquad	\frac{c^1{}_1}{\sqrt{2}}, \qquad \sqrt{2}c^1{}_1,
\end{equation*}
for $\psi_0,\cdots,\psi_4$ respectively.

We discretize the spatial directions by choosing equi-distant points in the 2-dimensional interval $[0.25,1.25]\times[0,\pi]$. As boundary conditions, we shoot in a gravitational quadrupole $(l=2)$ wave from each boundary by choosing the free data $q_0$ for $\psi_0$ on the right boundary and $q_4$ for $\psi_4$ on the left boundary as
\begin{gather*} 
q_0(s,\theta) =
	\begin{cases} 
	    2\sqrt{\frac{2\pi}{15}}\;{}_2Y_{20}(\theta)\sin^8(4{\pi s}),& s\leq\frac14 \\
	    0 & s>\frac14
	\end{cases}, \\
q_4(s,\theta) =
	\begin{cases} 
	    2\sqrt{\frac{2\pi}{15}}\;{}_2Y_{20}(\theta)\sin^8(4{\pi s}),& s\leq\frac14 \\
	    0 & s>\frac14
	\end{cases},
\end{gather*}
with the spin-weighted spherical harmonic ${}_2Y_{20}(\theta)\propto \sin^2{\theta}$ when written in the usual polar coordinates. We adopt our boundary treatment outlined in section~\ref{sec:ibvp} which fixes the boundary conditions for $\psi_1$ on the right boundary and $\psi_3$ on the left boundary. Hence we have fixed all the free data on the boundaries. We evolve up until $s=1$ with a $\theta$-resolution of $32$, $\rho$-resolutions of $\{25,50,100,200,400\}$ and use time-steps of $0.5\,\Delta\rho$. We find that this $\theta$-resolution is enough to represent all the functions in our system on the spheres $s=\text{constant}$, $\rho=\text{constant}$ at machine precision.

\begin{figure}[ht]
\centering
\subfigure[$s=0.02$]{
\includegraphics[scale=0.35]{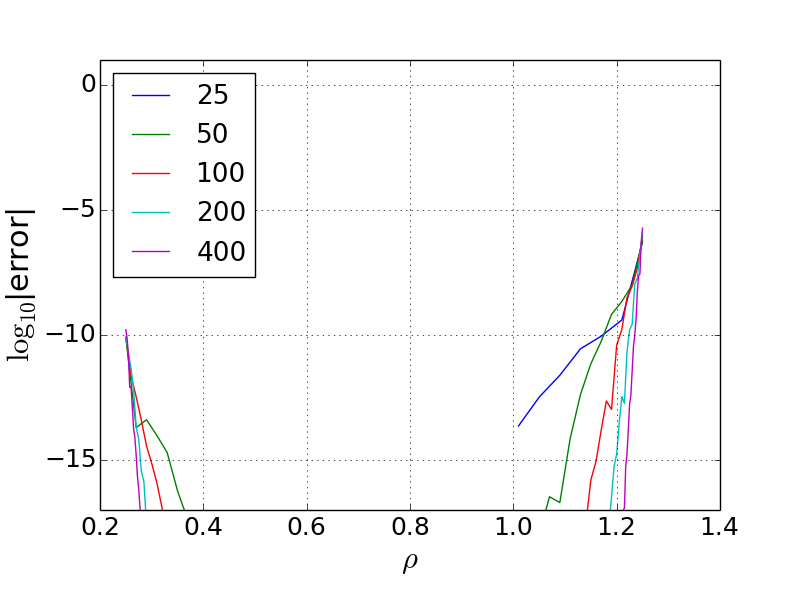}}
\subfigure[$s=0.42$]{
\includegraphics[scale=0.35]{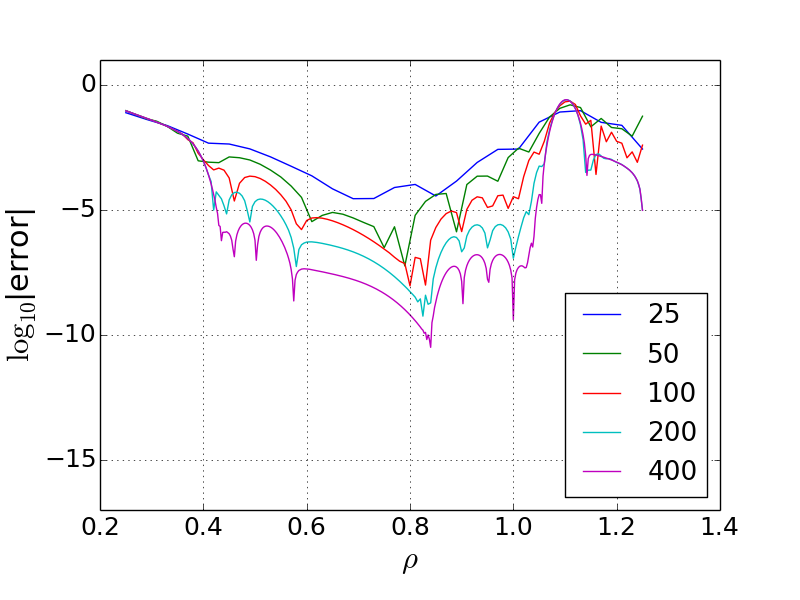}}
\subfigure[$s=0.62$]{
\includegraphics[scale=0.35]{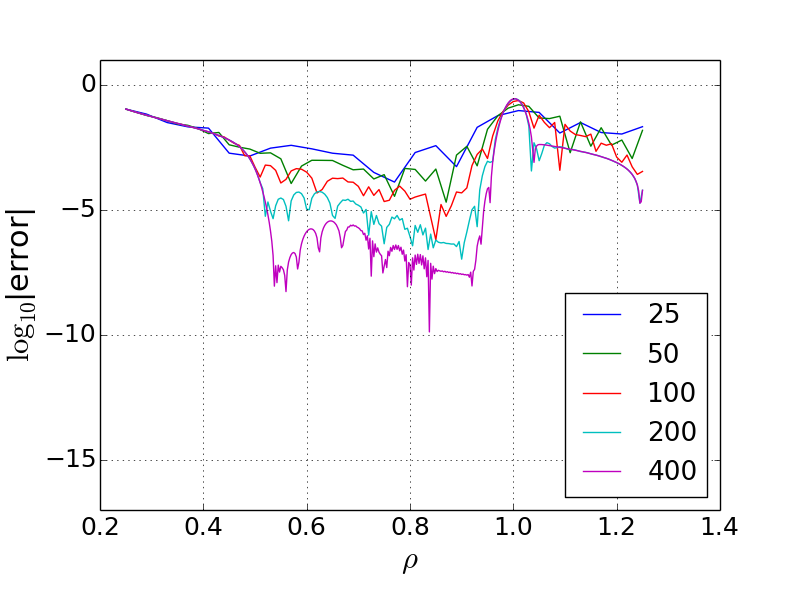}}
\subfigure[$s=0.98$]{
\includegraphics[scale=0.35]{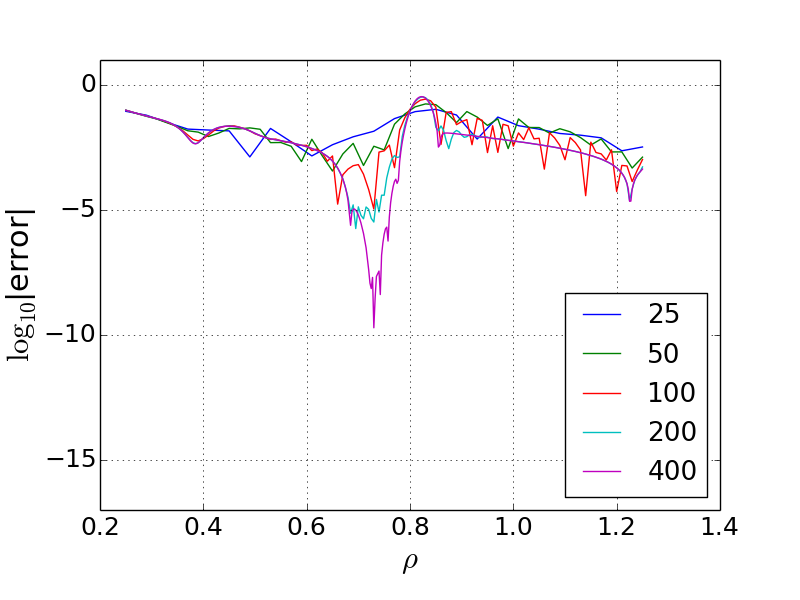}}
\caption{\label{fig:2waveswrongBC} A sequence of convergence tests at $s=\text{constant},\; \theta=\frac{\pi}{2}$ for a component of the $\psi_{ABCD}$ constraint $G_{AB}$ with increasing $\rho$-resolution for the case of two gravitational waves with Minkowski initial data using boundary conditions that do not kill subsidiary modes. Here error refers to the difference between the constraint and zero. As we expected, the constraint does not converge to zero everywhere.}
\end{figure}

First we look at the case without employing the subsidiary-mode-killing boundary treatment, to confirm that constraint violating modes are indeed propagated in from the boundary. We use the maximally dissipative boundary conditions but choose the free data (i.e.\ the ``$q$'') for $\psi_1$ on the right and $\psi_3$ on the left to be zero. Looking at the system variables on the slices $s=\text{constant},\; \theta=\frac{\pi}{2}$ in FIG.~\ref{fig:2waveswrongBC} we see that they converge everywhere with increasing resolution at the correct order of four. This means that the IBVP for the evolution system alone is well-posed. However, the constraint does \emph{not} converge to zero everywhere. The constraint violating modes should propagate in a time-like manner i.e., with less than the speed of light and this is clearly the case: with each snapshot the region where the constraint converges to zero becomes smaller while near the boundaries the constraint mode seems to converge to a non-zero limit. This indicates that the constraint is not satisfied.

\begin{figure}[ht]
\centering
\subfigure[$s=0.02$]{
\includegraphics[scale=0.35]{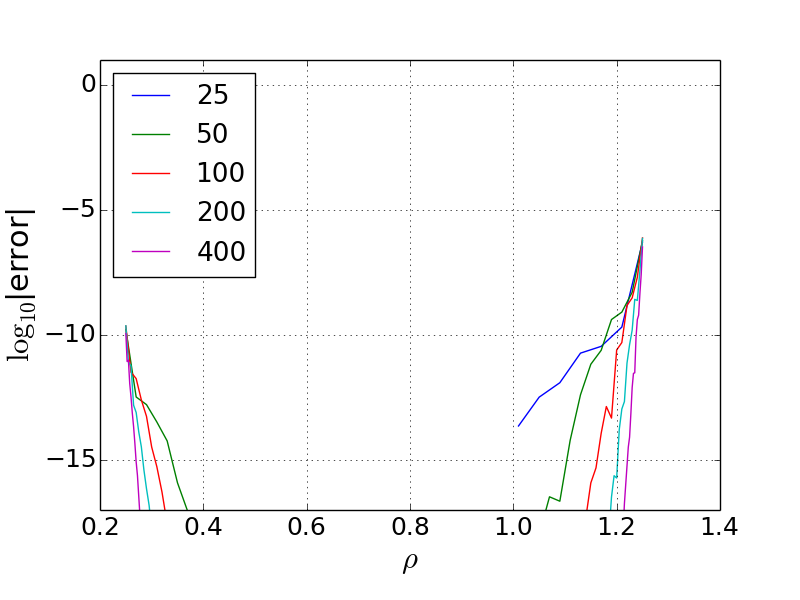}}
\subfigure[$s=0.42$]{
\includegraphics[scale=0.35]{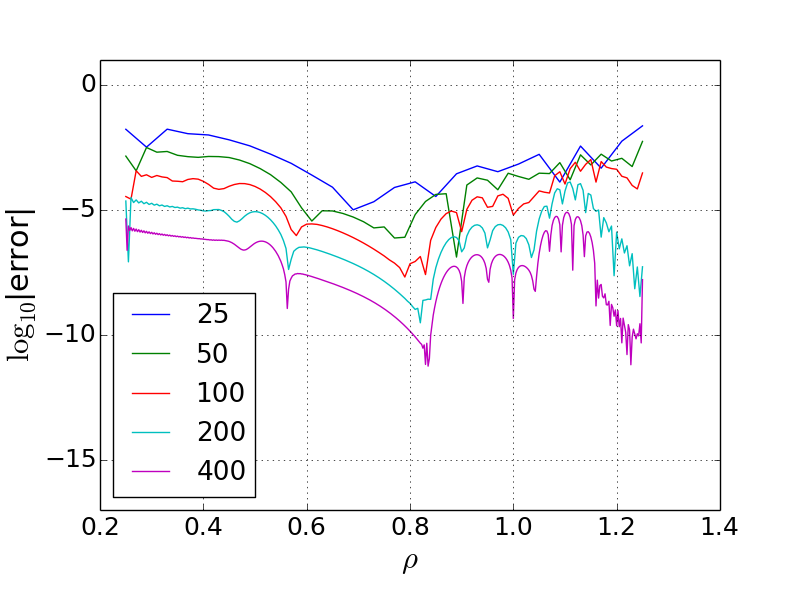}}
\subfigure[$s=0.62$]{
\includegraphics[scale=0.35]{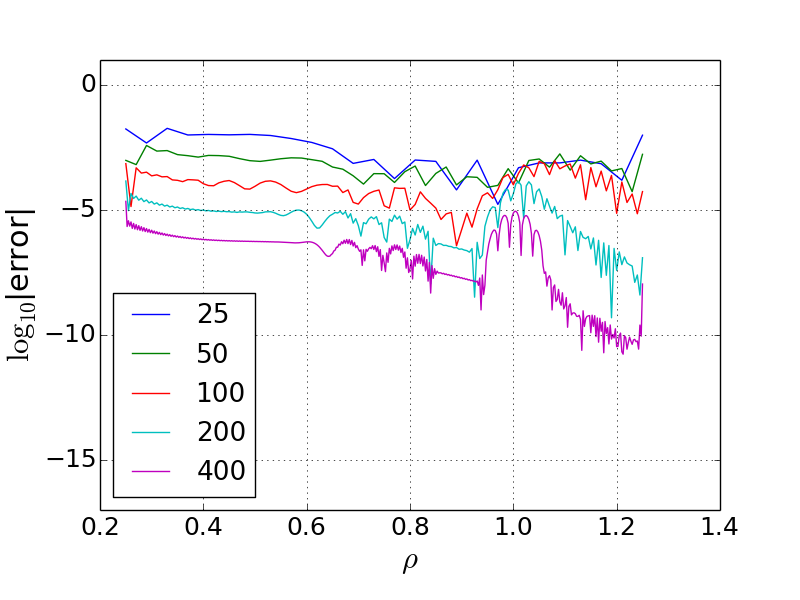}}
\subfigure[$s=0.98$]{
\includegraphics[scale=0.35]{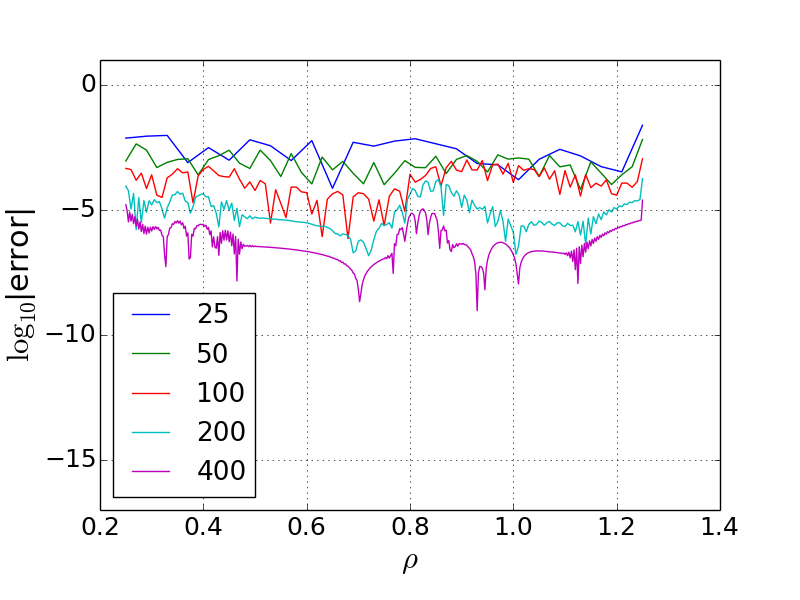}}
\caption{\label{fig:2wavescorrectBC} A sequence of convergence tests at $s=\text{constant},\; \theta=\frac{\pi}{2}$ for a component of the $\psi_{ABCD}$ constraint $G_{AB}$ with increasing $\rho$-resolution for the case of two gravitational waves with Minkowski initial data using boundary conditions that kill subsidiary modes. Here error refers to the difference between the constraint and zero. The constraint now converges to zero.}
\end{figure}

Next, we test the boundary treatment which supposedly kills the incoming constraint violating mode. In FIG.~\ref{fig:2wavescorrectBC} we display the same constraint as in FIG.~\ref{fig:2waveswrongBC} except with our constraint preserving boundary conditions. One can see that there is no longer a constraint violation propagating inward from the boundary and the constraints converge across the entire grid at the correct order.
Analogous plots are seen in all the other constraints and also for different choices of $\theta=$constant. Although this was a very simple case, the premise of our boundary treatment method has been verified.

% First simple test for IBVP framework where characteristic modes don't change sign.
% Shows framework works for simplest case.

\subsection{\label{subsec:schwarzschildIBVP}The IBVP for non-linear gravitational perturbations of Schwarzschild space-time}
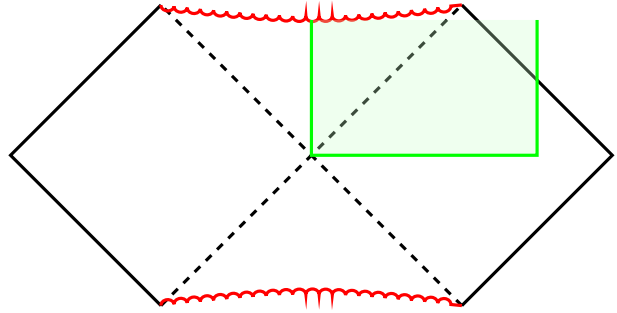
\begin{figure}[ht]
\centering
\begin{tikzpicture}[very thick, decoration = {bent,amplitude=-5}]
  \draw[dashed] (2,-2) -- (0,0)--(2,2);
  \draw (2,2)--(4,0) -- (2,-2);
  \draw[dashed] (-2,-2) -- (0,0)--(-2,2);
  \draw (-2,-2)--(-4,0) -- (-2,2);
  \draw[red,decoration={bumps,amplitude=-2},decorate] (-2,2) .. controls (0,1.8) .. (2,2);
  \draw[red,decoration={bumps,amplitude=2},decorate] (-2,-2)  .. controls (0,-1.8) .. (2,-2);
  \draw[draw  = green, very thick, fill=green!20!white, fill opacity=0.3] 
                          (0,1.8) -- (0,0) -- (3,0) -- (3,1.8);
\end{tikzpicture}
\caption{Schematic setup of the IBVP for Schwarzschild space-time perturbed by a gravitational wave. The location of the computational domain is indicated by the shaded area inside the Kruskal extension. The initial hyper-surface is a finite piece of the $T=0$ hyper-surface starting at the cross-over. It reaches up to the singularity and to null-infinity. This picture is perturbed by pumping in a gravitational wave from the outer (right) boundary.}
\label{fig:ourIBVPSchwarzschild}
\end{figure}

In the previous section we used simple initial data and fixed the gauge freedom appropriately so that we could test our framework for the simple case of $c^0_{AB}(s,\rho,\theta)=0$, i.e.\ the spatial frame vectors remain tangential to the $s=\text{constant}$ hyper-surfaces, in analogy to the Friedrich-Nagy gauge. However, this is a very special case and we want to explore the more general case of the non-vanishing $c^0_{AB}$ as well. We choose Schwarzschild space-time in isotropic coordinates as the test case and solve the associated IBVP with axi-symmetric boundary conditions as in the previous section. The IBVP setup is shown in FIG.~\ref{fig:ourIBVPSchwarzschild}.

It has been shown by Friedrich~\cite{friedrich2003conformal} that there exists a specific choice of initial data for the CGG that globally covers the Schwarzschild-Kruskal space-time smoothly and without degeneracy. This involves writing the Schwarzschild metric in isotropic coordinates and choosing appropriate initial data for the conformal geodesics. In this section we present how we (following~\cite{friedrich2003conformal}) set up the GCFE system to investigate null and time-like infinity of Schwarzschild space-time. We then numerically evolve the resulting initial data and see how the system behaves near the conformal boundary and the singularity.

The Schwarzschild metric written in isotropic coordinates is
\begin{equation}\label{bheq:1}
  \tg = \Big{(}\frac{1 - \frac{m}{2\rho}}{1 + \frac{m}{2\rho}}\Big{)}^2\text{d}t^2 - \Big{(}1 + \frac{m}{2\rho}\Big{)}^4\Big{[}\text{d}\rho^2 + \rho^2\Big{(}\text{d}\theta^2 + \sin^2{\theta}\text{d}\phi^2\Big{)}\Big{]}.
\end{equation}
We now take the hyper-surface $t=0$ as our initial hyper-surface $\Sigma_0$ and compute the corresponding initial data induced on it by the Schwarzschild metric. We also need to choose initial data $\ul{H}$, $\ul{Z}$ and $\ul{\Theta}$ so that we can compute the 1-form $h_a$ and conformal factor $\Theta$. Friedrich makes the choice
\[
	\ul{\Theta} = \frac{1}{r^2} = \frac{\rho^2}{(\rho + \frac{m}{2})^4},
\]
which does not compactify the initial surface in the $\rho$ direction. Next, the 1-form is fixed initially by setting $\ul{f}_a=0$ so that $\ul{h}_a=\nabla_a\ul{\Theta}$. This also determines $\ul{H}$ and we choose $\ul{Z}=0$ to fix the remaining freedom. Using \eqref{eq:10} and \eqref{eq:5}, this gives the explicit expression for $\Theta$ as
\begin{equation}\label{bheq:6}
	\Theta = \frac{\rho^2}{(\rho + \frac{m}{2})^4} - s^2\Big{(}\frac{\rho - \frac{m}{2}}{\rho + \frac{m}{2}}\Big{)}^2,
\end{equation}
and $h_a$, whose spinor representation we write in decomposed form as $h_{AB}+\frac12\eps_{AB}h$
\begin{equation}
\begin{gathered}
	h_{0} = h_{2} = 0, \\
	h_1 = -\sqrt{2}\rho\frac{\rho - \frac{m}{2}}{(\rho + \frac{m}{2})^3}, \qquad
	h = -2\sqrt{2}s\Big{(}\frac{\rho - \frac{m}{2}}{\rho + \frac{m}{2}}\Big{)}^2.
\end{gathered}\label{bheq:7}
\end{equation}
The final step is to prescribe the extrinsic curvature $K_{ABCD}$ which vanishes initially since the initial hyper-surface is time-symmetric.

From the initial metric chosen as \eqref{bheq:1} evaluated at $t=0$, expressions for $\Theta$ \eqref{bheq:6}, $h_{AB}$ \eqref{bheq:7} and $\ul{K}_{ABCD}=0$ we find the remaining non-vanishing initial data to be
\begin{gather*}
	R = \frac{\rho}{(\rho + \frac{m}{2})^2}, \qquad c^1{}_1 = \frac{(\rho + \frac{m}{2})^2}{\sqrt{2}},  \qquad
	\gamma_{20} = \hat{\gamma}_{01} = \frac{(\rho + \frac{m}{2})(\rho - \frac{m}{2})}{\sqrt{2}\;\rho}, \\[12pt]
	P_{101} = P_{110} = \frac{m(\rho + \frac{m}{2})^2}{\rho}, \qquad \psi_2 = -\frac{m(\rho + \frac{m}{2})^6}{\rho^3}.
\end{gather*}
A convenient function to know during the evolution is the original Schwarzschild radius $r(s,\rho)$. In spherical symmetry the radius of the 2-spheres in the GCFE system is $R$ and hence we can relate it the Schwarzschild radius by
\[
	r(s,\rho) = \frac{R}{\Theta}.
\]
This is very useful as it will tell us where the event horizon is located and if or when we end up at the curvature singularity. This is exactly true in spherical symmetry, i.e., for the unperturbed space-time but it will also be approximately correct in the perturbed case.

We choose $m=0.5$ and choose the spatial extent of the computational domain as the 2-dimensional interval $[0.25,1.25]\times[0,\pi]$ as in the previous section. Finally, we need to specify boundary conditions. The left boundary starts at the cross-over surface at $\rho=m/2=\frac14$. The Kruskal extension has a reflection isometry $\rho \mapsto m^2/4\rho$ which fixes the hyper-surface $\rho=m/2$. We impose this reflection symmetry as boundary condition on the left boundary. This implies the following conditions for the ingoing components of the rescaled Weyl spinor
\[ 
	\psi_3(s,0.25,\theta) = -\bar{\psi}_1(s,0.25,\theta),\qquad \psi_4(s,0.25,\theta) = \bar{\psi}_0(s,0.25,\theta).
\]
There will be no constraint violation propagating in from the inner boundary as we have not violated the constraints there, it will remain as Schwarzschild space-time until the gravitational wave coming in from the right boundary reaches it, which does not happen in our simulations.

On the right boundary we implement our constraint preserving boundary treatment. This is done by first choosing the free wave profile $q_0$ for $\psi_0$ to be
\[ 
	q_0(s,\theta) =
	\begin{cases} 
	    4a\sqrt{\frac{2\pi}{15}}\;{}_2Y_{20}(\theta)\sin^8(4{\pi s})& s\leq\frac14 \\
	    0 & s>\frac14
	\end{cases},
\]
where $a$ is a fixed constant representing the amplitude of the wave. The other ingoing mode is $\psi_1$. Its boundary value is determined to kill the ingoing constraint violating mode via our resolution.

Now we have all that is needed to start evolving the system. We note that we use a $\theta$-resolution of $64$ so that even at late times of the simulation our fields are well represented in the spin-weighted spherical harmonic basis. We use $\rho$-resolutions of $\{25,50,100,200,400\}$ which gives us ample data to check the propagation of the constraints. We evolve up to $s=1.22$ which incorporates $\mathscr{I}^+$ into the simulation, which is reached on the right boundary at around $s=0.83$. Simulations are also performed without our boundary treatment where the appropriate $q_i,\;i\neq0$ are set to zero. This will allow us to contrast the before and after of our boundary treatment and emphasize the problem that we resolve.

It is worth noting that we could change the boundary treatment when the right boundary has passed beyond $\mathscr{I}^+$. As the constraint violating modes are time-like, they can never propagate through $\mathscr{I}^+$ from outside and hence there is no need to kill them any longer. However, while this is true analytically we must be cautious with this kind of argument since the numerical propagation of constraint violation can be faster. This is a purely numerical artefact which will shrink as the $\rho$-resolution is increased. 
% \begin{figure}[ht!]
% \begin{center}
% \begin{tikzpicture}[very thick, decoration = {bent,amplitude=-5}]
% 	\draw[red,decoration={bumps,amplitude=-2},decorate] (0,4) -- (4,4);
% 	\draw[dashed] (4,4) -- (0,0);
%  	\draw (4,4) -- (8,0); 	

%  	\draw[green] (0,0) -- (6,0);

%  	\draw[green] plot [smooth] coordinates {(7,2) (6.5,2.5) (5.5,3) (2,3.5)};
%  	\draw[green] (0,3.5) -- (2,3.5);

%  	\draw[blue] (6,0) -- (6,0.2) .. controls (6.2,0.8) .. (7,2);
%  	\draw[blue] (5,0) -- (5,0.2) .. controls (5.2,1.5) .. (6,2.83);
%  	\draw[blue] (4,0) -- (4,0.2) .. controls (4,1.5) .. (4.5,3.15);
%  	\draw[blue] (3,0) -- (3,0.2) .. controls (3,1) .. (3.3,3.32);
%  	\draw[blue] (2,0) -- (2,0.2) .. controls (2.1,3.46) .. (2.1,3.48);
%  	\draw[blue] (1,0) -- (1,0.2).. controls (1.07,3.49) .. (1.07,3.49);
%  	\draw[blue] (0,0) -- (0,3.5);
% \end{tikzpicture}
% \caption{The geometrical ramifications of $g^{00}$ going to zero. The green lines are $s=\text{constant}$ surfaces while the blue lines are conformal geodesics. The conformal geodesics remain time-like while part of the later $s=\text{constant}$ surface transitions from space-like to null.}
% \label{fig:bendingofboundary}
% \end{center}
% \end{figure}

\begin{figure}[ht]
\centering
\subfigure[$s=0.02$]{
\includegraphics[scale=0.35]{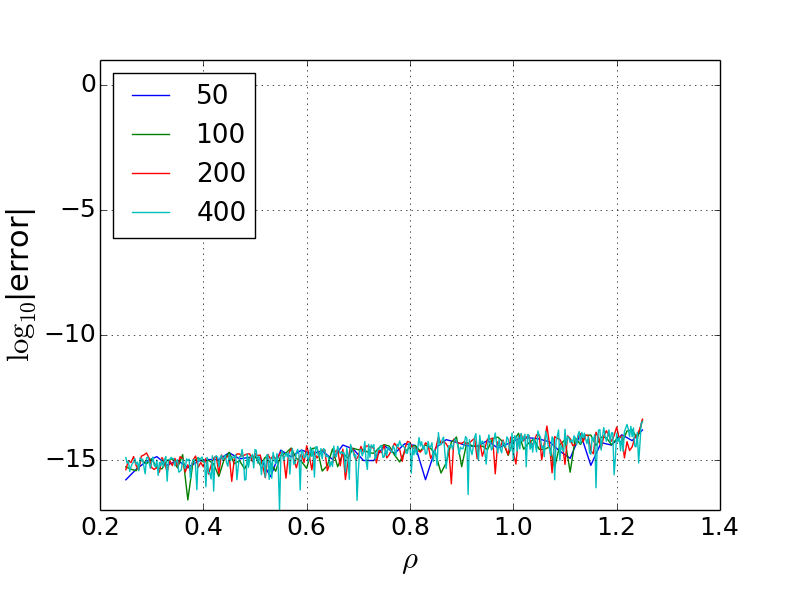}}
\subfigure[$s=0.42$]{
\includegraphics[scale=0.35]{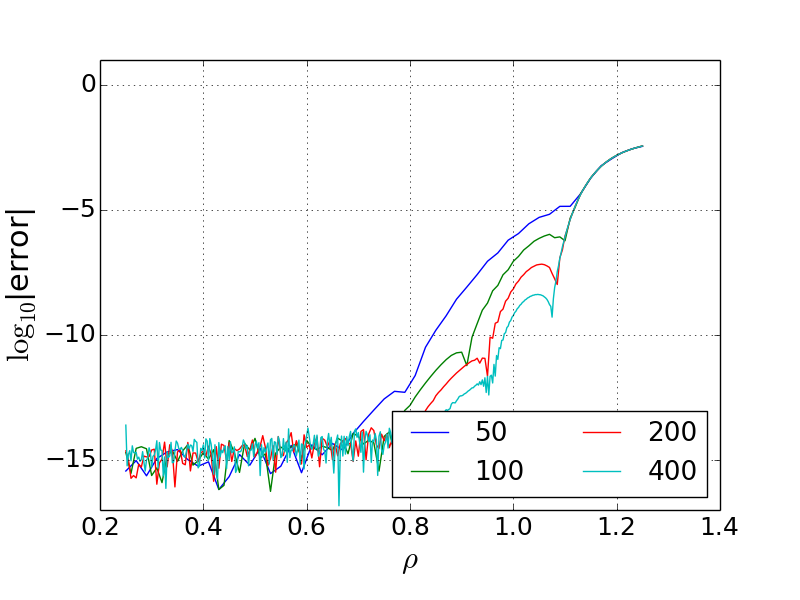}}
\subfigure[$s=0.82$]{
\includegraphics[scale=0.35]{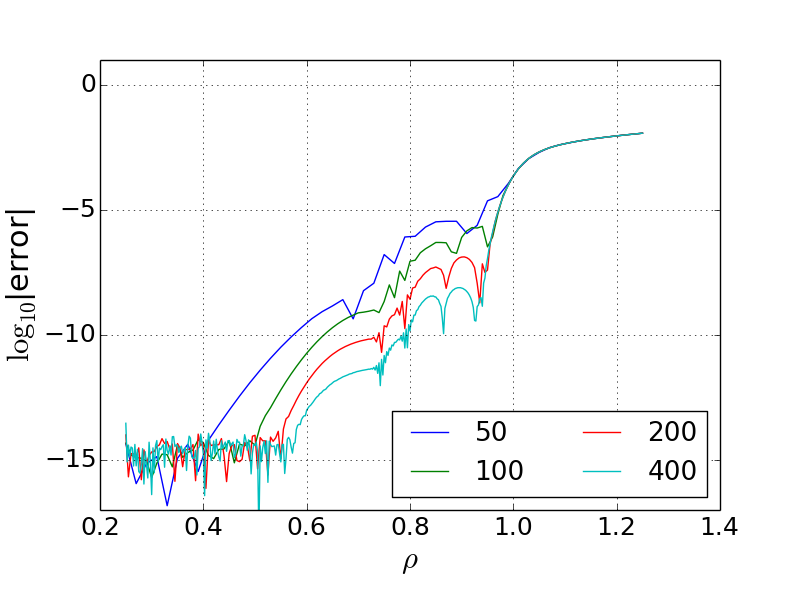}}
\subfigure[$s=1.2$]{
\includegraphics[scale=0.35]{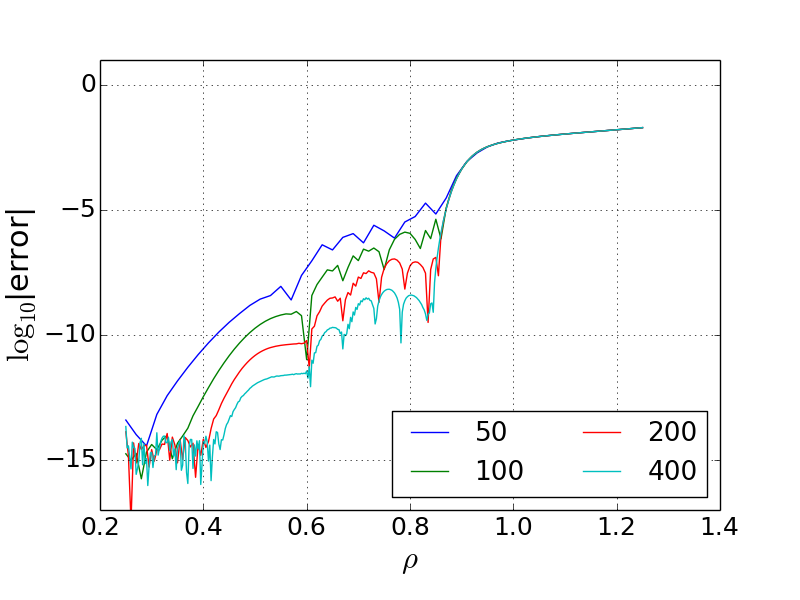}}
\caption{\label{fig:pertschwarzschildconstrconv1} A sequence of convergence tests for a component of the $\psi_{ABCD}$ constraint $G_{AB}$ with non-constraint-preserving boundary conditions imposed computed on $s=\text{constant},\;\theta=\frac{\pi}{2}$ slices with increasing $\rho$-resolution. Here error refers to the difference between the constraint and zero. As expected, the constraint does not converge to zero on the entire domain.}
\end{figure}

We first present a convergence plot for the case of the simple, but non-constraint-preserving, choice of setting the free boundary data (except $q_0$) to zero. FIG.~\ref{fig:pertschwarzschildconstrconv1} displays the convergence plots for a component of the $\psi_{ABCD}$ constraint $G_{AB}$ at $s=\text{constant},\;\theta=\frac{\pi}{2}$ slices. One clearly sees that there is a mode propagating in from the right boundary that stops the constraints from converging to machine precision. This happens not just to this constraint, but to all in the constraints in our system.

\begin{figure}[ht]
\centering
\subfigure[$s=0.02$]{
\includegraphics[scale=0.35]{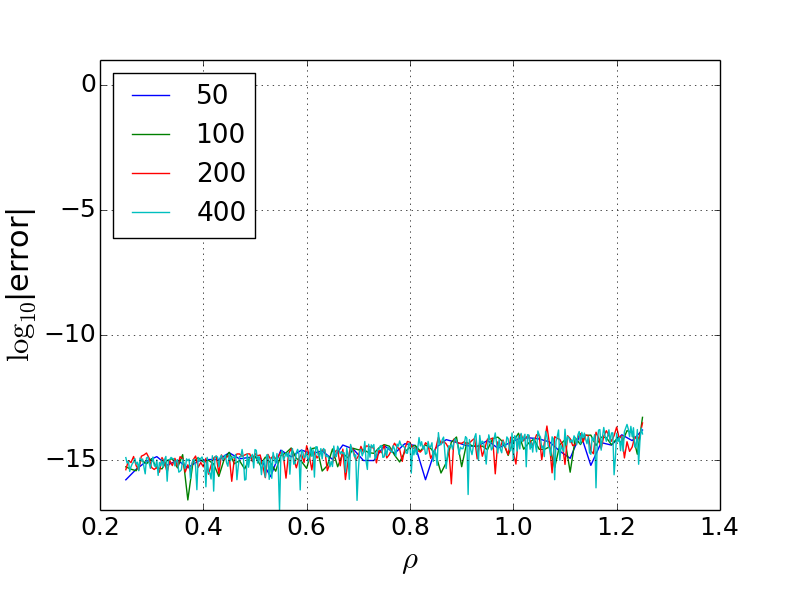}}
\subfigure[$s=0.42$]{
\includegraphics[scale=0.35]{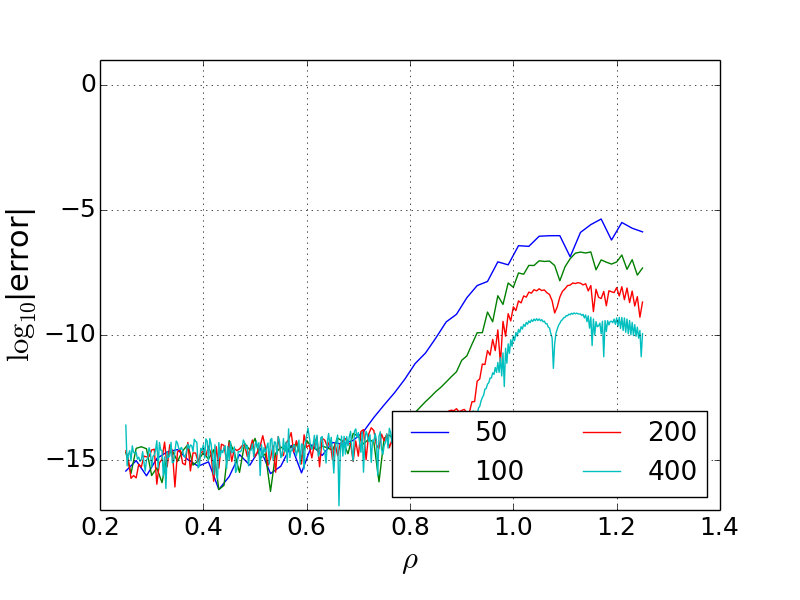}}
\subfigure[$s=0.82$]{
\includegraphics[scale=0.35]{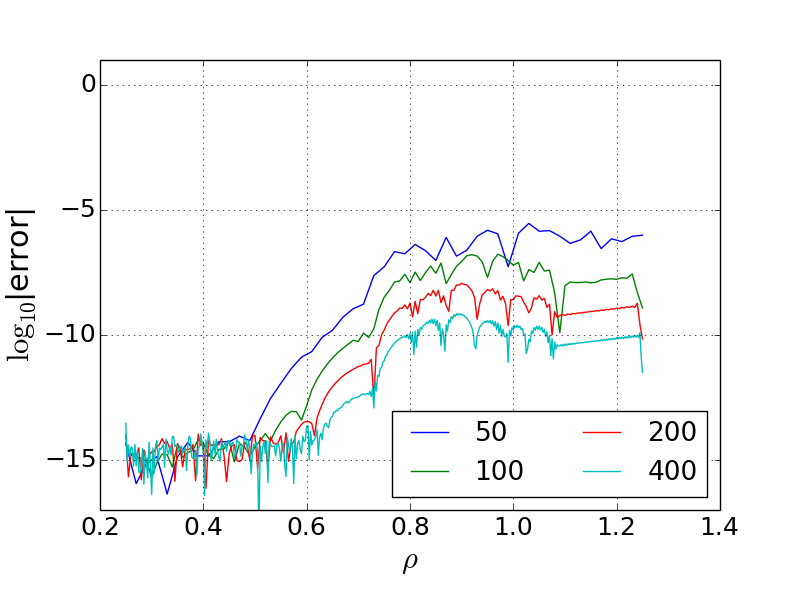}}
\subfigure[$s=1.2$]{
\includegraphics[scale=0.35]{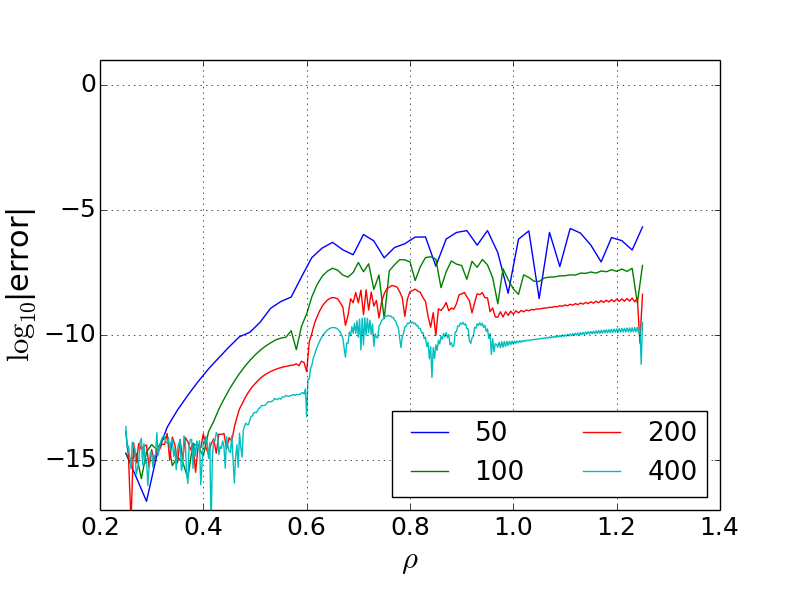}}
\caption{\label{fig:pertschwarzschildconstrconv2} A sequence of convergence tests for a component of the $\psi_{ABCD}$ constraint $G_{AB}$ with constraint-preserving boundary conditions imposed computed on $s=\text{constant},\;\theta=\frac{\pi}{2}$ slices with increasing $\rho$-resolution. Here error refers to the difference between the constraint and zero. As we expected, the constraint now converges to zero everywhere.}
\end{figure}

Now we contrast these plots to the analogous ones that implement our boundary treatment, shown in FIG.~\ref{fig:pertschwarzschildconstrconv2}. Immediately one sees that these convergence plots are exceedingly better than the previous ones. We get convergence toward machine precision at the correct order and in the process have moved the constraints around $1\times10^{8}$ closer to this. Looking at other constraints in the system we see that the problem has been overcome in all of them and this is seen to be the case for different choices of $\theta$. 

Thus we have shown that without our boundary treatment, constraint violating modes of the subsidiary system propagate into the computational domain and prevent the constraints from converging to zero. In the constraint violating case we see again that the evolution system is well-posed since the constraint violation converges to non-zero value.
% We have also shown in detail that killing all but one of the ingoing constraint modes still leads to destruction of convergence.
Thus the most general case of our boundary treatment has been shown to be numerically viable.

\begin{figure*}[htp]
  \centering \subfigure[$\theta=\frac{\pi}{4}$]{\includegraphics[scale=0.38]{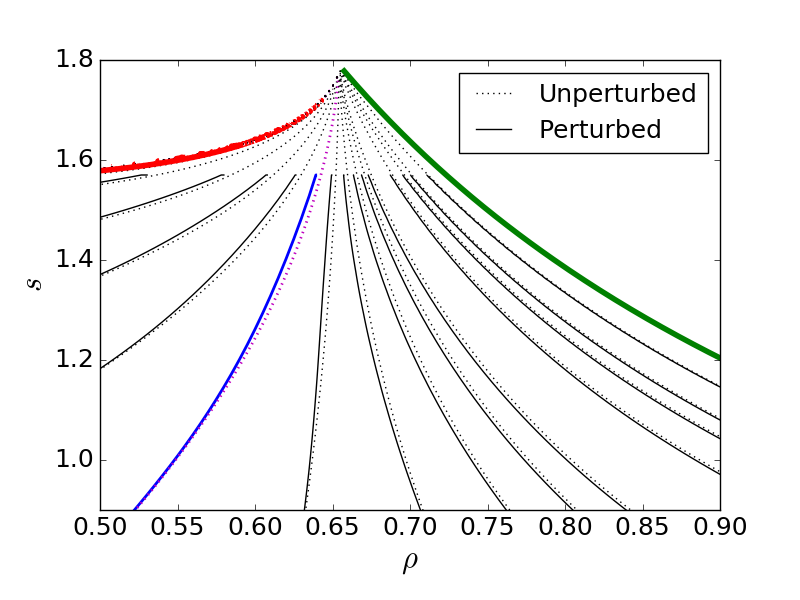}}\quad \subfigure[$\theta=\frac{\pi}{2}$]{\includegraphics[scale=0.38]{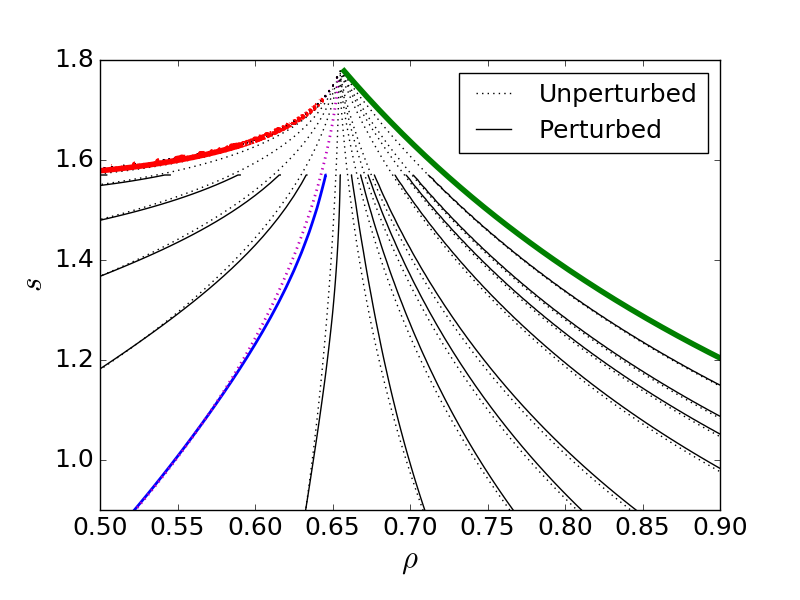}}
  \label{fig:rcontours}
  \caption{Two contour plots of the Schwarzschild radius $r$ in both the perturbed and unperturbed space-times. The curves closest to the bottom left are the $r=1$ curves (the event horizon in the unperturbed space-time). The outermost curves are curves very close to $r=0$ (left) and $\mathscr{I}^+$ (right). Note that in the unperturbed case we can get further than the perturbed case due to the system consisting only of ODEs, whence approaching $r=0$ does not destroy the simulation.}
\end{figure*}

To end this section, it is interesting to see how the ``approximate'' Schwarzschild radius $r=R/\Theta$ in the perturbed case differs to that of the unperturbed case, see Figure \ref{fig:rcontours}. One can see that there is in fact a deviation that gets larger as the simulation progresses. This deviation is large enough to be discerned by the eye alone, which tells us that the location of the event horizon in the perturbed case may have actually moved a significant distance. These plots also showcase the anisotropic nature of the gravitational radiation, with spheres being stretched in one direction and squeezed in the other. This can be deduced by noticing the ``perturbed'' Schwarzschild radius is larger or smaller than the corresponding unperturbed one, depending on which part of the sphere is considered.

% \begin{figure}[ht]
% \makebox[\linewidth][c]{
% \begin{subfigure}[b]{0.55\textwidth}
% \centering
% \includegraphics[width=.95\textwidth]{images/rschwarzscontour_closeup_piover4.png}
% \caption{}
% \end{subfigure}%
% \begin{subfigure}[b]{0.55\textwidth}
% \centering
% \includegraphics[width=.95\textwidth]{images/rschwarzscontour_closeup_piover2.png}
% \caption{$\theta=\frac{\pi}{2}$}
% \end{subfigure}
% }
% \caption{Two contour plots of the Schwarzschild radius $r$ in both the perturbed and unperturbed space-times. The curves closest to the bottom left are the $r=1$ curves (the event horizon in the unperturbed space-time). The outermost curves are curves very close to $r=0$ (left) and $\mathscr{I}^+$ (right). Note that in the unperturbed case we can get further than the perturbed case due to the system consisting only of ODEs, whence approaching $r=0$ does not destroy the simulation.}
% \label{fig:rcontours}
% \end{figure}

\section{\label{sec:summary}Summary}

% Discuss future work and the scope of the framework.
In this paper we discussed a numerical implementation of the GCFE in the space-spinor formalism. Subsequently the manifold topology was restricted to be of the form $M_2\times\mathbb{S}^2$ so that the $\eth$-calculus could be utilized. The resulting system was checked for correctness by an analytic comparison to exact solutions as well as a range of numerical tests.

Most of the evolution equations in the system are ODEs in the sense that they involve only the time derivative $\del_s$ with the notable exception of the subsystem for the gravitational spinor $\psi_{ABCD}$, which is symmetric hyperbolic. We discussed the characteristics of both that system and the part of the subsidiary system concerned with the propagation of its constraints. Using maximally dissipative boundary conditions we proposed a practical way of imposing boundary conditions which are physically reasonable in the sense that there is one freely specifiable degree of freedom and no incoming constraint violating modes. This was found to be numerically stable in several different settings, indicating that the IBVP for this formulation of the conformal Einstein equations is well posed.

% Once our numerics and system were checked for correctness, we moved on to investigate the viability of our constraint preserving boundary treatment. This was shown to work for the case of gravitational perturbations of Minkowski space-time in $2+1$ dimensions where the gauge was fixed in such a way that it was essentially the standard Gauss gauge in physical space-time. For this special case the spatial frame vectors remained tangential to the $s=\text{constant}$ surfaces and the frame vector normal to the boundaries remained normal, in analogy to the Friedrich-Nagy gauge. Our framework was then proven to work for the case of gravitational perturbations of Schwarzschild space-time, also in $2+1$ dimensions, where the spatial frame vectors titled out of the $s=\text{constant}$ surfaces. The system was evolved up to and beyond $\mathscr{I}^+$. Thus we have successfully developed a procedure for setting up an IBVP for the GCFE where coordinates $x^0=\text{constant}$, $x^1=\text{constant}$ are topological 2-spheres and where the boundaries are given by time-like conformal geodesics. It is numerically well-posed in the sense that constraints propagate.

This newly developed framework will give us the ability to investigate a range of problems that involve the global structure and global properties of space-times. In a forthcoming paper we will report on studies of the characteristic ringing behaviour of the Schwarzschild space-time under gravitational perturbations and discuss the global issues of wave signal readout on $\scri^+$ and the mass-loss due to gravitational radiation. After this problem, the most obvious next step to take is to generalize this setup to the Kerr space-time. As it is still not clear whether this space-time is stable under general perturbations, we could investigate this question from a global perspective. Investigating this problem does not require any fundamental changes to the setup for the Schwarzschild space-time, only a new initial data set is needed.

There is also the possibility of investigating the conditions required on an asymptotically flat initial data set so that the resulting vacuum solution has a regular null infinity. Friedrich \cite{friedrich2007static} has restricted the problem to how initial data is chosen on the blowup of the point $i^0$ to a 2-sphere. He has conjectured that the necessary condition for a regular null infinity is that the initial data near null infinity are those induced by asymptotically conformally stationary space-times. This still remains as just a conjecture and hence it would be intriguing to probe this question numerically by evolving sets of initial data that do and do not satisfy the necessary conditions of the conjecture. There have already been numerical studies of linearly perturbed space-times which incorporate space-like infinity, see for example \cite{beyer2012numerical,Doulis:2017kg,Doulis:2013wp}, while in~\cite{frauendiener2014fully,frauendiener2016fully} we have studied simpler systems which show similar behaviours near space-like infinity.

Another open problem within the scope of this approach is that of the stability of Anti-de Sitter space-time. First brought to attention by Bizo\'{n} and Rostworowski in 2012~\cite{bizon2011weakly}, it was found that the space-time is non-linearly unstable under a particular class of perturbations. Since then time-stable periodic solutions have been discovered~\cite{maliborski2013time,balasubramanian2014holographic,green2015islands}.  The main issue to address in this context is the fact that a conformal geodesic on the Einstein cylinder ``stalls'' in the sense that it does only cover a finite interval of conformal time even for an infinite range of its parameter. This means that we need to implement a reparametrization in order to ``reset'' its parameter and to continue the evolution.

On the more mathematical side of things it would be desirable to have a rigorous proof that our empirical boundary treatment does in fact lead to a well-posed IBVP.

\appendix

\section{\label{subsec:spacespinors}The space-spinor formalism}

Due to its compact form, the calculus of space-spinors \cite{sommers1980space} is a very enticing formalism in which to write the GCFE. We will use definitions similar to that used in the series of papers by Frauendiener \cite{frauendiener1998numerical1,frauendiener1998numerical2,frauendiener2000numerical} throughout.

First, we define a time-like vector field $t^a$, normalized with respect to the conformal metric $g$ by $t^at_a=2$. Writing $t^a$ in terms of spinors, one immediately finds the relationship
\[
	t_{AA'}t_{B}{}^{A'} = t_{A'(A}t_{B)}{}^{A'} + \frac12t_{CA'}t^{CA'}\eps_{AB} = \eps_{AB},
\]
which justifies the choice of normalization constant. One can now use this vector field as a map from the complex spin-space $\bar{S}^{A'}$ onto the spin-space $S^A$, i.e.\ we can convert primed indices to unprimed ones. This map is given by
\begin{equation*}
	\alpha_{A'}\mapsto t_{A}{}^{A'}\alpha_{A'} =: \alpha_A.
\end{equation*}
For example, a spinor $\alpha_a=\alpha_{AA'}$ can be mapped to $\alpha_{AB} := t_B{}^{A'}\alpha_{AA'}$. The result can be decomposed into two terms

\[
	\alpha_{AB} = \alpha_{(AB)} + \frac12\eps_{AB}\alpha_E{}^E, \qquad \text{with} \quad \alpha_E{}^E = \alpha_{EA'}t^{EA'}.
\]
This shows us that the trace term corresponds to the part of the spinor that has values in the direction of $t^{AA'}$. Thus finding irreducible decompositions of space-spinors is the same as performing a 3+1 splitting. This is incredibly useful for deriving evolution and constraint equations.

We also define a complex conjugation map on the unprimed spin-space via
\begin{equation*}
	\alpha_A\mapsto\hat\alpha_A:=t_A{}^{A'}\bar{\alpha}_{A'}.
\end{equation*}
This map has the property that for a spinor of rank $n$ we obtain
\[
	\widehat{\hat\alpha}_{AB\cdots D} = (-1)^n\alpha_{AB\cdots D}.
\]
We define a rank-2 spinor as real iff it is equal to the negative of its complex conjugate, in accordance with the reality of the SL(2,$\mathbb{C}$) spinors. Since for any two real rank-2 spinors $\alpha_{AB}$ and $\beta_{CD}$ their outer product should also be real, a rank-4 spinor is real if it is equal to its complex conjugate.

We can now split the covariant derivative $\nabla$ into spatial and temporal parts using the mapping $t_B{}^{A'}\nabla_{AA'}$ and its subsequent decomposition. This gives us two new derivative operators
\begin{equation*}
D = t^{AA'}\nabla_{AA'}, \quad D_{AB} = t_{(A}{}^{B'}\nabla_{B)B'} \implies 
\nabla_{AA'} = \frac12 t_{AA'} D - t^{B}{}_{A'} D_{AB}.
\end{equation*}
Two fundamental spinor fields can now be defined as the derivatives of $t^a$ with respect to these new derivative operators. We have
\begin{equation*}
K_{CD} := t_D{}^{C'}D t_{CC'}, \qquad 
K_{ABCD} := t_D{}^{C'}D_{AB} t_{CC'}.
\end{equation*}
Geometrically, the spinor field $K_{AB}$ corresponds to the acceleration vector of $t^a$ while $K_{ABCD}$ is related to the geometry of the distribution defined by vectors $V^a$ that satisfy $V^at_a=0$ and for a time-like vector field $t^a$ satisfying the hyper-surface orthogonal property, it corresponds to the extrinsic curvature. They have the reality properties
\begin{equation*}
	\hat{K}_{AB} = -K_{AB},\qquad \hat{K}_{ABCD} = K_{ABCD}.
\end{equation*}

Note that these new derivatives operators are real in the sense that they map real spinors to real spinors, but they do not commute with our definition of complex conjugation, i.e.\
\[
	D\hat\alpha_C = D (\bar\alpha_{A'}t_C{}^{A'}) =  t_C{}^{A'} D \bar\alpha_{A'} + \bar\alpha_{A'}\,D t_C{}^{A'}= \widehat{D\alpha_C} + \hat\alpha_{A}\,K_C{}^{A}.
\]
A similar equation holds for $K_{ABCD}$. Hence we introduce new derivative operators
\begin{equation*}
	\del \alpha_C = D\alpha_C - \frac12 K_C{}^D \alpha_D, \qquad 
	\del_{AB} \alpha_C = D_{AB}\alpha_C - \frac12 K_{ABC}{}^D \alpha_D,
\end{equation*}
adjusted to commute with complex conjugation. We now have
\begin{equation*}
	\del_{AB}\hat\alpha_C = - \widehat{\del_{AB}\alpha_C},\qquad \del\hat\alpha_C = \widehat{\del\alpha_C}.
\end{equation*}
Note that the correction term $K_{AB}$ or $K_{ABCD}$ corrects only one index. So replacing the action of $D$ or $D_{AB}$ on a rank-$n$ spinor with $\del$ or $\del_{AB}$ respectively, result in $n$ correction terms. For example,
\[
	\del \alpha_{AB} = D\alpha_{AB} - \frac12 K_A{}^C \alpha_{CB} - \frac12 K_B{}^C \alpha_{AC}.
\]
We will denote the spin-frame spinors by $o^A$ and $\iota^A$ along with their primed counterparts.

\section{The complete subsidiary system}
\label{sec:compl-subs-syst}

It is convenient in the calculation of the subsidiary equations to decompose the fields $K_{ABCD}$, $P_{ABCD}$ and $\gamma_{ABC}$ into irreducible pieces. The irreducible decompositions of these fields can be written as
\begin{gather}
	\gamma_{ABC} = \tgam_{ABC} + 2\tgam_{(A}\eps_{B)C},\\[12pt]
	K_{ABCD} = \tK_{ABCD} + \frac12\Big{(}\eps_{A(C}\tK_{D)B} + \eps_{B(C}\tK_{D)A}\Big{)} - \frac13\tK\eps_{A(C}\eps_{D)B}, \\[12pt]
	P_{ABCD} = \tP_{ABCD} + \frac12\Big{(}\eps_{A(C}\tP_{1D)B} + \eps_{B(C}\tP_{1D)A}\Big{)} + \frac12\eps_{CD}\tP_{2AB} - \frac13\tP\eps_{A(C}\eps_{D)B},
\end{gather}
where the new spinor quantities are totally symmetric. It is also useful to decompose the constraint $U_{ABCD} = \tU_{ABCD} + \frac12\tU_{AB}\eps_{CD}$, where $\tU_{ABCD}=\tU_{(AB)(CD)}$ and $\tU_{AB}=\tU_{(AB)}$.

Applying the above procedure to the most general form of the constraints \eqref{constreqs} we find the subsidiary system to be
\begin{subequations}\label{subsideqs}
	\begin{align}
	\del Z^0_{AB} &= -\frac{1}{\sqrt{2}}T_{AB} + \frac14C^{0CD}Z_{ABCD}, \label{subsideq1}\\
	\del Z^i_{AB} &= \frac14C^{iCD}Z_{ABCD}, \label{subsideq2}\\
	\del G_{AB} &= \del_{(A}{}^CG_{B)C} - \frac32K_{(A}{}^CG_{B)C} + \frac12\tK_{(A}{}^CG_{B)C} \nonumber \\
		&+ \frac12G^{CD}\tK_{ABCD} - \frac43\tK G_{AB} - T^{CD}\psi_{ABCD} - \frac12Z_{(A}{}^{CDE}\psi_{B)CDE} - 2Z^{ECD}{}_E\psi_{ABCD}, \label{subsideq3}\\
	\del T_{AB} &= T^{CD}\tK_{ABCD} - \tK_{(A}{}^CT_{B)C} - \frac23\tK T_{AB} + U_{AB} + \frac14K^{CD}Z_{ABCD}, \label{subsideq4}\\
	\del Z_{ABCD} &= \Theta\eps_{AC}\Big{(}G_{BD} - \hG_{BD}\Big{)} + \Theta\eps_{BD}\Big{(}G_{AC} - \hG_{AC}\Big{)} - 4U_{ABCD} \nonumber \\
        &- \tK Z_{ABCD} + Z_{ABE(C}\tK_{D)}{}^E - \tK_{(A}{}^EZ_{B)ECD} + \tK_{ABEF}Z^{EF}{}_{CD} - \tK_{CDEF}Z_{AB}{}^{EF}, \label{subsideq5}\\
	\del J_{ABC} &= \tK_{D(A}J_{B)}{}^D{}_C + J^{DE}{}_C\tK_{ABDE} - \frac23\tK J_{ABC} 
		+ \frac14\Theta o_B\Big{(}G_{AC} + \hG_{AC}\Big{)} + \frac12\tU_{AB}o_C \nonumber \\
		&+ \frac14Z_{ABEF}K_C{}^Fo^E + \hgam^DZ_{ABCD} - \frac12\hgam_{CDE}Z_{AB}{}^{DE} 
		- \frac14\Theta\eps_{AC}\Big{(}G_{BD} + \hG_{BD}\Big{)}o^D, \label{subsideq6}\\
	\del \tU_{AB} &= \frac12h_A{}^C\Big{(}G_{BC} - \hG_{BC}\Big{)} + \frac12h^C{}_B\Big{(}G_{AC} - \hG_{AC}\Big{)} - \frac23\tK \tU_{AB} \nonumber \\
		&- \tK_{(A}{}^C\tU_{B)C}+ \tK_{ABCD}\tU^{CD} + \frac12Z_{ABCD}\tP_2^{CD}, \label{subsideq7}\\
	\del \tU_{ABCD} &= -\frac12h_{B(C}G_{D)A} - \frac12\hG_{A(C}h_{D)B} - \frac23\tK\tU_{ABCD} - \tK_{(A}{}^EU_{B)ECD}
		+ \tK_{ABEF}U^{EF}{}_{CD} \nonumber \\
		&- \frac16\tP Z_{ABCD} + \frac12Z_{ABE(C}\tP_{1D)}{}^E - \frac12\tP_{CDEF}Z_{AB}{}^{EF}
		+ \frac12\eps_{A(C}h^E{}_{D)}G_{BE} + \frac12\eps_{A(C}h_{D)}{}^E\hG_{BE}. \label{subsideq8}
	\end{align}
\end{subequations}

% If you have acknowledgments, this puts in the proper section head.
\begin{acknowledgments}
The authors are grateful to H. Friedrich for discussions. JF wishes to thank the Department of Mathematics at the University of Oslo for hospitality while this article was written. Funding for this visit was provided by the European Research Council through the FP7-IDEAS-ERC Starting Grant scheme, Project No. 278011 STUCCOFIELDS. The authors wish to acknowledge the contribution of NeSI high-performance computing facilities to the results of this research.
\end{acknowledgments}

% Create the reference section using BibTeX:
%\bibliography{paper,papers}
\bibliography{paper}

\end{document}